\documentclass[aps,prl,notitlepage,twocolumn,superscriptaddress,longbibliography,nofootinbib]{revtex4-1}

\usepackage[T1]{fontenc} 
\usepackage{lmodern}
\usepackage{amsmath,amsthm,amssymb,amsfonts}
\usepackage{scalerel}
\usepackage{mathtools}
\usepackage{graphicx}
\usepackage{subfigure}
\usepackage[normalem]{ulem}
\usepackage{tabularx}
\usepackage[colorlinks = true,
            linkcolor = blue,
            urlcolor  = blue,
            citecolor = blue,
            anchorcolor = blue]{hyperref}
\usepackage{mathrsfs}
\usepackage{bbold}
\usepackage{units}
\allowdisplaybreaks
\usepackage{bm}
\usepackage{braket}
\usepackage{makecell}
\usepackage[nameinlink]{cleveref} 
\usepackage{upgreek}
\usepackage{blindtext}
\usepackage{verbatim}
\usepackage{algorithm}
\usepackage{balance}
\usepackage[noend]{algpseudocode}
\makeatother

\usepackage[dvipsnames]{xcolor}
\usepackage{bbm}
\usepackage[bb=boondox]{mathalfa}
\usepackage{array}
\usepackage{multirow}
\usepackage{tabularx}
\usepackage{float}
\usepackage{dcolumn}
\usepackage{slashed}
\usepackage{braket}
\usepackage{verbatim}
\usepackage{multirow}
\usepackage{resizegather}
\usepackage{titlesec}
\usepackage[toc,page]{appendix}
\usepackage[export]{adjustbox}
\newcommand{\appendixhead}%
{\centering\textbf{\huge Appendices}
\vspace{0.25in}}
\def\lf{\left\lfloor}   
\def\rf{\right\rfloor}
\definecolor{teal}{RGB}{0, 158, 115} 
\definecolor{morange}{RGB}{255, 127, 0}

\newcommand{\kfim}{\mathrm{KFIM}\,}
\newcommand{\dd}{\mathcal{D}}

\DeclareMathOperator{\Tr}{Tr}

\begin{document}

\title{Extreme value statistics and eigenstate thermalization in kicked quantum chaotic spin-$1/2$ chains}

\author{Tanay Pathak\,\,\href{https://orcid.org/0000-0003-0419-2583}
{\includegraphics[scale=0.05]{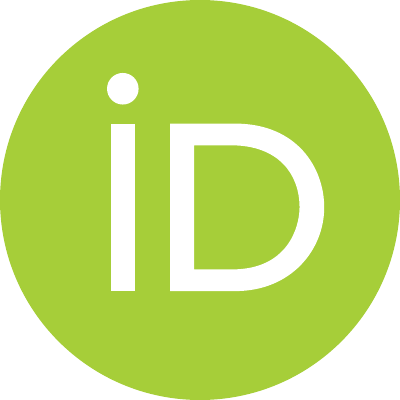}}}
\email{pathak.tanay.4s@kyoto-u.ac.jp}
\affiliation{Center for Gravitational Physics and Quantum Information, Yukawa Institute for Theoretical Physics,\\ Kyoto University, Kitashirakawa Oiwakecho, Sakyo-ku, Kyoto 606-8502, Japan}
\affiliation{Department of Physics, Kyoto University, Kitashirakawa Oiwakecho, Sakyo-ku, Kyoto 606-8502, Japan}
\author{Masaki Tezuka\,\,\href{https://orcid.org/0000-0001-7877-0839}
{\includegraphics[scale=0.05]{orcidid.pdf}}\,}
\email{tezuka@scphys.kyoto-u.ac.jp}
\affiliation{Department of Physics, Kyoto University, Kitashirakawa Oiwakecho, Sakyo-ku, Kyoto 606-8502, Japan}

\begin{abstract}
It is often expected (and assumed) for a quantum chaotic system that the presence of correlated eigenvalues implies that all the other properties as dictated by random matrix theory are satisfied. We demonstrate using the spin-$1/2$ kicked field Ising model that this is not necessarily true. We study the properties of eigenvalues of the reduced density matrix for this model, which constitutes the entanglement spectrum. It is shown that the largest eigenvalue does not follow the expected Tracy--Widom distribution even for the large system sizes considered. The distribution instead follows the extreme value distribution of Weibull type. Furthermore, we also show that such deviations do not lead to drastic change in the thermalization property of this system by showing that the model satisfies the diagonal and off-diagonal eigenstate thermalization hypothesis. Finally, we study the spin-spin autocorrelation function and numerically show that it has the characteristic behavior for chaotic systems: it decreases exponentially and saturates to a value at late time that decreases with system size.
\end{abstract}
~~~~~~~~~~~~Report Number: YITP-25-75

\maketitle

\emph{Introduction}: 
Random matrix theory (RMT) and quantum chaos are intimately connected. The presence of RMT behavior is considered a hallmark of quantum chaos. Single particle quantum systems are chaotic if there is a level repulsion among the eigenvalues as is stated by the Bohigas--Giannoni--Schmit (BGS) conjecture \cite{PhysRevLett.52.1,bohigas1984spectral,casati1980connection,berry1985semiclassical}. The behavior of eigenvalues and their correlations are well described by random matrix theory (RMT)\cite{mehta2004random,forrester2010log}. For integrable systems we have the Berry--Tabor conjecture \cite{berry1977level,marklof2001berry}, which states that for a quantum system whose classical counterpart is integrable, the eigenvalues in quantum systems do not have any correlation, they show level clustering and in the semi-classical limit behave statistically similar to a random process. These two conjectures are assumed to hold for many-body quantum systems as well. The presence of level repulsions in many-body quantum systems is often taken to be a signature of quantum chaos \cite{Hsu_1993,PhysRevLett.70.497,Jacquod_1997,PhysRevB.66.052416,PhysRevLett.106.050405,Gharibyan:2018jrp} and it is often assumed (and expected) that it also implies that all the other properties of RMT are followed as well. It is also important to emphasise that when we say RMT features we usually consider the prototypical random matrix ensembles such as the Gaussian or circular ensemble. It is entirely possible to introduce an appropriate random matrix ensemble such that certain properties of these ensembles match those of Gaussian or circular ensemble, yet there might still be properties that they capture due to \emph{extra} structure introduced. 

However, the situation for many-body quantum systems is not straightforward as the Hamiltonians of many well-known many-body quantum systems contain only local interactions and are highly sparse unlike random matrices, which are dense and contain matrix elements of $k= 1, 2, \cdots, N$-body type\cite{Guhr:1997ve}. Thus it is not clear if it is reasonable to expect RMT behavior in these systems. Moreover, deviations from RMT in many-body systems have been observed at the edge of the spectrum \cite{RevModPhys.82.277}, in the eigenvector distribution \cite{PhysRevLett.126.121602}, fractal dimensions \cite{PhysRevE.100.032117,PhysRevLett.126.150601} and also for the entanglement entropy \cite{Pausch:2020pwc,Vidmar:2017uux,Vidmar:2017pak,Haque_2022,Huang:2017std,huang2021universal}. The situation, however, is different for the case of kicked systems. These systems are more scrambling in the sense that the time evolution operator is less local unlike the case of isolated many-body systems. Their dynamics can be described by an effective Hamiltonian that has interactions in all possible ranges and orders \cite{Herrmann:2023hdj} and thus it is expected that the RMT features are more closely followed by these systems.

In this work, by taking a prototypical model of quantum chaos, the kicked field Ising model (KFIM), we explore the question: \emph{Are all the properties of RMT observed in a model of many body quantum chaos and to what extent the deviations from the RMT behavior one should still expect the thermalisation properties of the system of be in accordance to the expectations?}
KFIM has been well studied as a minimal model of quantum chaos \cite{Bertini:2018wlu, PhysRevE.65.036208,Bertini:2018fbz,Braun:2019kgf,Flack:2020ybm}. It has also been recently explored for its deviation from random matrix behavior in its entanglement statistics \cite{Herrmann:2024wat}.

\emph{Kicked field Ising model:}
The kicked field Ising model for $L$ spins with periodic boundary condition has the following time evolution operator
\begin{align}
    U_{\mathrm{KFIM}} &= e^{-i H_{z}} e^{-i H_{x}}, \quad \text{with} \nonumber \\
    H_{z} &= J \sum_{i=1}^{L} \sigma_{i}^{z}\sigma_{i+1}^{z} + \sum_{i=1}^{L} h_{i} \sigma_{i}^{z},\nonumber \\
    H_{x} &= b \sum_{i=1}^{L} \sigma_{i}^{x},
\end{align}
where $\sigma_{i}^{x}$ and $\sigma_{i}^{z}$ are the standard Pauli matrices. The $H_{z}$ part contains the nearest neighbor coupling of the $z-$component of spin with strength $J$ and a position-dependent longitudinal field $\{h_{i}\}$. The $H_{x}$ part denotes the driving Hamiltonian with a periodic kick and kicking strength $b$. For our present study we consider the self-dual KFIM, $J=b= \pi/4$, where the model is known to be maximally chaotic \cite{akila2016particle,Kos:2017zjh,Bertini:2018wlu,Lerose:2020fhd}. The $h_{i}$'s are randomly chosen from the standard normal distribution. The suitable random matrix ensemble associated with the model is BDI in Cartan's classification \cite{duenez2004random,Flack:2020ybm,Braun:2019kgf}. However, following \cite{Herrmann:2023hdj,Herrmann:2024wat}, we use the circular orthogonal ensemble (COE) as a reference for simplicity.

\emph{Marchenko--Pastur law and extreme value distributions}
The randomness (RMT behavior) and ergodicity are intertwined phenomena. The Hamiltonian (or unitary time evolution operator) of quantum chaotic system can be modelled using Gaussian (or unitary) random matrices. Thus for such systems the eigenstates can be assumed to be purely random, and the intensity of the eigenstates follows the Porter--Thomas law \cite{porter1965statistical, mehta2004random}. We consider a bi-partition of Hilbert space $\mathcal{H} = \mathcal{H}_{1} \otimes \mathcal{H}_{2}$, with $\text{dim}( \mathcal{H}_{1}) = \dd_{1}$, $\text{dim}( \mathcal{H}_{2}) = \dd_{2}$, $\dd_{1} \leq \dd_{2}$ and $\text{dim}( \mathcal{H}) = \dd_{1}\dd_{2}$. If $\Psi$ is a state belonging to this Hilbert space then we have the following Schmidt decomposition of this state with respect to the basis states $\ket{j}_{1} \in \mathcal{H}_{1}$ and $\ket{j}_{2} \in \mathcal{H}_{2}$,
\begin{equation}
    \Psi = \sum_{j_{1}=1}^{\dd_{1}} \sum_{j_{2}=1}^{\dd_{2}} c_{j_{1}j_{2}} \ket{j_{1}} \otimes \ket{j_{2}} = \sum_{j'=1}^{\dd_1} \sqrt{\lambda_{j'}} \ket{j'}_{1}\ket{j'}_{2},
\end{equation}
where $\lambda_{j'}$ are the Schmidt coefficients and $\ket{j'}_{1}$, $\ket{j'}_{2}$ are obtained as linear combinations of $\{|j_1\rangle\}, \{|j_2\rangle\}$ with the coefficients obtained as the elements of the normalized eigenvectors of the matrix $CC^\dagger, C^{\dagger}C$. 
The reduced density matrix of the subsystems is then given by $\rho_{1}= C C^{\dagger}$ and $\rho_{2}= (C^{\dagger}C)^{T}$, where $C$ is the coefficient matrix whose elements are $c_{ij}$. $\lambda_{i}$'s are further important in the study of entanglement properties, and the largest $\lambda_{i}$ carry the largest weight in determining the entanglement properties \cite{Laflorencie_2016}. Thus, it is important to understand the features of the maximum $\lambda_{i}$ and how well they follow the RMT.
Furthermore, the structure of the reduced density matrix corresponding to these ergodic eigenstates belongs to the trace-restricted Wishart ensemble (Wishart matrices with unit trace) $\equiv \frac{W}{\Tr(W)}$, where $W$ is the Wishart matrix. From hereon we refer to them as Wishart matrices to avoid clumsy wordings. The joint probability distribution of the eigenvalues of Wishart matrices is known exactly \cite{mehta2004random}. The average density of the eigenvalues follow the Marchenko--Pastur law \cite{mehta2004random}, which for the special case when $\dd_{1}= \dd_{2}= \dd$ is given by 
\begin{equation}\label{eqn:mplaw}
   P(\tilde{e})= \frac{1}{2 \pi} \sqrt{\frac{4- \tilde{e}}{\tilde{e}}}, \quad 0\leq \tilde{e} \leq 4,
\end{equation}
where $\tilde{e}= \dd \lambda_{i}$ are the rescaled Schmidt coefficients. 

For numerical purposes, we consider $L= 8(10^6)$, $9(500000)$, $10 (500000)$, $11(500000)$, $12(300000)$, $13(179000)$, $14(101600)$, $15(72400)$, $16(51200)$, $18(40960)$, numbers in the bracket denoting the total number of eigenvectors considered over many realizations. Since the eigenvalues of the unitary matrices lie on the unit circle, we obtained a few eigenvalues and eigenvectors near the phase $\phi= \frac{\pi}{2}$, each for the COE random matrix and the KFIM. We use the shift-invert method for system sizes $L \leq 12$ and polynomially filtered exact diagonalization (POLFED) \cite{Luitz_2021,Sierant:2022xtl} for $L \geq 13$ (see Supplemental Material \cite{supp} for further details on the algorithm used). For the study of distribution of Schmidt coefficients we only consider even $L$ and trace out first $L/2$ spin for the case of KFIM or $2^{L/2}$-dimensional Hilbert space for other models, to form the reduced density matrix. Full diagonalization is done for the numerical calculations of the autocorrelation function for sizes $L= 8 (10000), 9 (5000), 10 (2000), 11 (1000), 12 (500), 13 (100),\\ 14 (50)$, where the number in brackets denotes the disorder realizations considered.

\begin{figure}[tbp]
    \centering
    \includegraphics[width=  \linewidth]{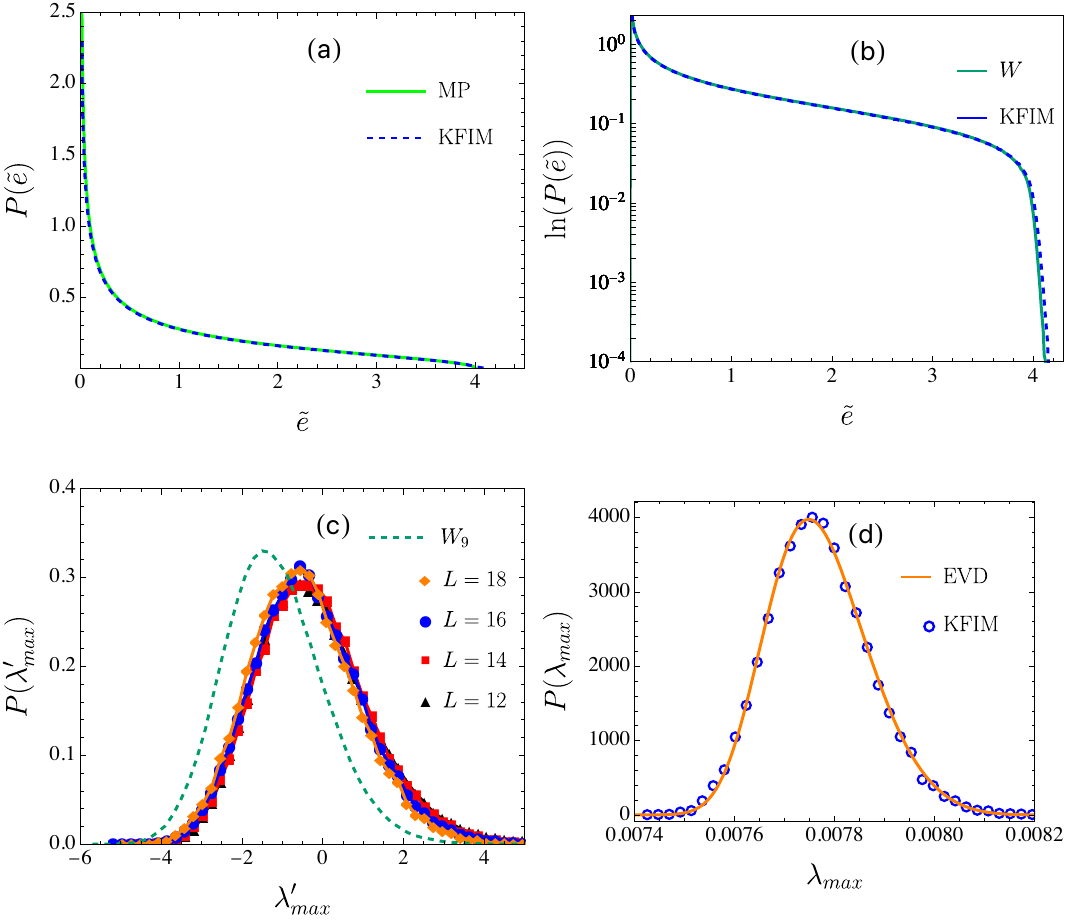}
    \caption{
    (a) Comparison of distribution of $\tilde{e}$ for KFIM ($L=18$) with Marchenko--Pastur distribution given by Eq. \eqref{eqn:mplaw}. (b) Comparison of distribution of $\tilde{e}$  obtained using $2^{9} \times 2^{9}$ dimensional Wishart matrices ($W_{9}$) with KFIM ($L=18$) in semi-logarithmic plot. (c) Distribution of $\lambda'_{\max}$ obtained for Wishart matrices ($W_9$ and KFIM for $L= 12, 14, 16, 18$. The $\lambda'_{\max}$ is obtained after suitable centering and rescaling the largest Schmidt coefficient using Eq. \eqref{eqn:cenres}. (d) Fitting of maximum Schmidt coefficient $\lambda_{\max}$, for KFIM ($L=18$), to extreme value distribution given by Eq. \eqref{eqn:evdg}.  The distribution given by Eq. \eqref{eqn:evdg} is suitably centered and rescaled with free parameters $\alpha$ and $\beta$ respectively.}\label{fig:mplaw}
\end{figure}

The result for the distribution of $\tilde{e}$, for KFIM with $L=18$ is shown in Fig. \ref{fig:mplaw}(a). However, the Marchenko--Pastur law is an asymptotic distribution, only realized fully for infinite-dimensional matrices. So to account for the finite size effects we instead numerically calculate the distribution using Wishart matrices of dimension $2^{9} \times 2^{9}$ and averaged over $10^{6}$ samples (see Supplemental Material \cite{supp} for further details on why this is justified). The results for KFIM are then compared in a semi-logarithmic scale as shown in Fig. \ref{fig:mplaw}(b). For the case of KFIM, near the tail, we observe deviations that decrease with $L$, which suggest that the properties of the largest eigenvalue might be different than anticipated.  It is also important to note that the Marchenko--Pastur distribution is a universal distribution for an ensemble of correlation matrices, independent of the exact distribution of matrix elements given it has a finite moment of sufficiently larger order \cite{tao2012random}. 

To further probe the tail of the distribution we study the properties of the maximum Schmidt coefficient. Its properties can be deduced using the eigenvalue properties of Wishart matrices. Typical maximum eigenvalue of Wishart matrices is given by \cite{johnstone2001} $ \lambda_\mathrm{max}= \mu_W+ \sigma_W \lambda'_\mathrm{max}$, where $\lambda'_\mathrm{max}$ is a random variable that characterizes the typical fluctuations and it follows the Tracy--Widom distribution with 
\begin{align}\label{eqn:cenres}
\mu_W &= \frac{\left(\sqrt{\dd-1}+\sqrt{\dd}\right)^2}{\dd^2}\sim \frac{4}{\dd}, \nonumber \\
\sigma_W &=\frac{\left(\sqrt{\dd-1}+\sqrt{\dd}\right) \sqrt[3]{\frac{1}{\sqrt{\dd}}+\frac{1}{\sqrt{\dd-1}}}}{\dd^2}    \sim 2^{\frac{4}{3}} \dd^{-\frac{5}{3}},
\end{align}
where $\sim$ implies the result for large dimension $\dd$. We then study the distribution of $\lambda'_\mathrm{max}= \frac{\lambda_\mathrm{max}- \mu(W)}{\sigma(W)}$ and the result asymptotically should be the Tracy--Widom distribution $F_{1}$. As stated before, to properly account for the finite size effect, we numerically obtain the distribution of the largest eigenvalue of Wishart matrices of dimension $2^{9} \times 2^{9}$ using $10^{6}$ realizations. The comparison of the thus obtained numerical distribution, with KFIM, is shown in Fig. \ref{fig:mplaw}(c). We observed that even for the largest $L=18$ there are still significant deviations from the result of Wishart matrices.   

Deviations from the Tracy-Widom distribution suggest the existence of different extreme value distributions in these systems. To verify this we use the Fisher--Tippett--Gnedenko theorem \cite{fisher1928limiting,gumbel1958,Majumdar_2020}. The theorem states that for a set of identical and independent random variables, $\{Z_{1},Z_{2},\cdots , Z_{N}\}$, with probability density given by $p(z)$, the probability density of the maximum of $Z_{i}$'s after suitable centering and rescaling should follow one of the three limiting distributions: Fisher--Tippett--Gumbel, Weibull and Fr\'echet. The resulting distribution depends on the tail of the density of $p(z)$, whether it is power law or faster than any power law and whether it is bounded or unbounded. The probability distribution function of the suitably centered and rescaled extreme (maximum) values $y\equiv(z-\mu)/\sigma$ is given as follows.
\begin{align}\label{eqn:evdg}
P(y ; \tilde{\xi})& = 
 \begin{dcases}
        \exp \left(-(1-\tilde{\xi} y)^{1 / \tilde{\xi}}\right)(1-\tilde{\xi} y)^{(1 / \tilde{\xi}-1)} & \tilde{\xi} \neq 0, \\
        \exp (-\exp (-y)) \exp (-y)& \tilde{\xi}=0.
 \end{dcases}
\end{align}
with $\tilde{\xi}$ being the shape parameter. $\tilde{\xi}= 0$, $\tilde{\xi}< 0$, and $\tilde{\xi}> 0$ imply Gumbel, Fr\'echet, and Weibull distributions respectively. These universal distributions are still valid in the presence of weak correlations \cite{leadbetter1988extremal}. For the case of KFIM we now study the distribution of $\tilde{\lambda}_{\max}= \frac{\lambda_{\max}- \alpha}{\beta}$, where $\alpha$ and $\beta$ are the free parameters to be determined from the fit along with $\tilde{\xi}$ in Eq. \eqref{eqn:evdg}. However, for numerical purposes, it is suitable to instead study $\lambda_{\max}$ and suitably recentre and rescale the distribution Eq. \eqref{eqn:evdg} using $x \rightarrow \frac{x -\alpha}{\beta}$. Doing this we obtain the best fit parameters: $\alpha = 7.77 \times 10^{-3} \pm 8.38 \times 10^{-7}, \beta= 9.40 \times 10^{-5} \pm 6.44 \times 10^{-7}, \tilde{\xi}= 0.17 \pm 0.011$, suggesting Weibull distribution. The numerical results for the same along with the best fit curve are shown in Fig. \ref{fig:mplaw} (d).

\emph{Approach towards Tracy--Widom distributions:} From Fig. \ref{fig:mplaw}(c), it is not clear if with increasing $L$ the distribution approach towards Tracy--Widom distribution or not. To quantify the approach towards the Tracy--Widom distribution we consider the approach of moments of numerically obtained distributions towards these distributions. However, even if the mean and the standard deviation of the two approach each other, it is possible that the distributions remain distinct \cite{Rodriguez-Nieva:2023err,Herrmann:2024wat}. So we instead consider the ratio $\mathcal{R}= \frac{\Delta_{\overline{\lambda'}_{\max}}}{\sigma_{\mathrm{TW}}}$, with $\Delta_{\overline{\lambda'}_{\max}}= | \overline{\lambda'}^{\kfim}_{\max}- \mu_{\mathrm{TW}}|$, where $\overline{\lambda'}^{\kfim}_{\max}$ denotes the mean value for KFIM and $\mu_{\mathrm{TW}}, \sigma_{\mathrm{TW}}$ denotes the mean and the variance respectively of the Tracy--Widom distribution. Intuitively, $\mathcal{R}$ measures how many standard deviations away the two distributions are from each other. 

\begin{figure}[ht]
    \centering
    \includegraphics[width=  \linewidth]{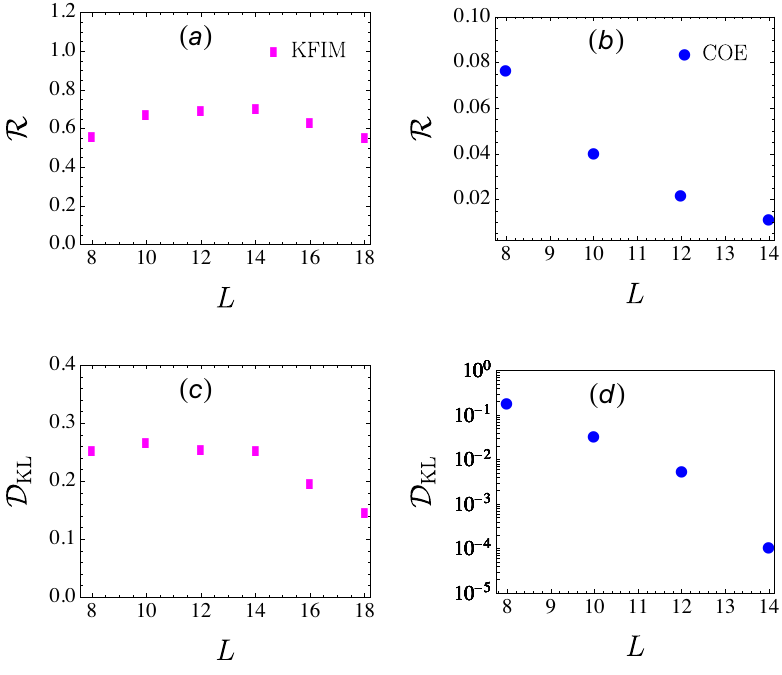}
    \caption{Value of $\mathcal{R}= \frac{\Delta_{\overline{\lambda'}_{\max}}}{\sigma_{\mathrm{TW}}}$ as a function of $L$ (a) for the KFIM, (b) for the COE.  KL divergence, $\mathcal{D}_{\mathrm{KL}}$ (c) for the KFIM. (d) for the COE. It can be observed that for the COE case the $\mathcal{R}$ as well as $\mathcal{D}_{\mathrm{KL}}$ is much lower as compared to KFIM for the corresponding value of $L$ and goes rapidly to zero. The $\mathcal{D}_{\mathrm{KL}}$ values for COE show a clear decreasing trend with $L$, while for KFIM a slight decreasing trend starts only after $L =14$.}\label{fig:momratioall}
\end{figure}

We expect that if $\mathcal{R}$ is decreasing (increasing) with $L$, the two distributions, the ones obtained for KFIM and the Tracy--Widom distribution, approach (deviate from) each other. If $\mathcal{R}$ has some small constant value, it means that the two distributions are close enough but never approach each other. For the case of KFIM, shown in Fig. \ref{fig:momratioall}(a) we do not observe a clear decreasing trend for $\mathcal{R}$ and find that for the largest $L (=18)$, $\mathcal{R}= 0.56$, indicating that the distributions approach quite close to each other but still remain distinct. On the other hand, for the COE case shown in Fig.~\ref{fig:momratioall}(bd), it decreases with $L$ thus showing convergence towards the Tracy--Widom distribution. It is also important to note that for the COE case, $\mathcal{R}$  has much lower value as compared to KFIM for the same system size thus indicating closeness even for small system sizes.

We also study the Kullback–Leibler (KL) divergence for these models. The KL divergence, $\mathcal{D}_{\mathrm{KL}}$, for distribution $Q(x)$, obtained using the experimental data, and $P(x)$, the theoretical distribution, is given as: 
\begin{equation}
    \mathcal{D}_{\mathrm{KL}}= \sum_{x} P(x) \log \left(\frac{P(x)}{Q(x)}\right) .
\end{equation}
$\mathcal{D}_{\mathrm{KL}}=0$ implies the two distribution are identical. We study the variation of  $\mathcal{D}_{\mathrm{KL}}$ with system size $L$ for KFIM and COE case. The results are shown in Fig. \ref{fig:momratioall} (c) and (d) respectively. We observe a (nearly) constant value of $\mathcal{D}_{\mathrm{KL}}$ till $L=14$ and then observe that it decreases slowly after that and reaches a value of $\mathcal{D}_{\mathrm{KL}}= 0.15$ for $L=18$. In contrast for the COE case even for the smallest value, $L=8$ the value of $\mathcal{D}_{\mathrm{KL}}= 0.18$ and then decreases rapidly, reaching $\mathcal{D}_{\mathrm{KL}} \approx 10^{-4}$ for $L=14$, showing rapid convergence towards the Tracy--Widom distribution. Due to the unclear trend for the KFIM case it is difficult to difficult to conclude concretely, with the system size available, if the Tracy--Widom distribution is really achieved asymptotically, i.e. in $L \rightarrow \infty$ limit.

\emph{Eigenstate Thermalisation Hypothesis and autocorrelation function:}
For a given unitary operator, we seek to study the (quasi)eigenstate thermalisation hypothesis, similar to the Hamiltonian eigenstates. In particular, we seek to study the expectation of an observable $\mathcal{O}$ in (quasi)eigenstate, $U \ket{n}= e^{i \phi}\ket{n}$. We study the following ETH ansatz: $\braket{n|\mathcal{O}|m}= \mathcal{O}(\bar{\phi}) \delta_{mn}+ \rho(\bar{\phi})^{-1/2}f(\bar{\phi},\omega)R_{mn}$.

\begin{figure}[htbp]
    \centering
    \includegraphics[width=  \linewidth]{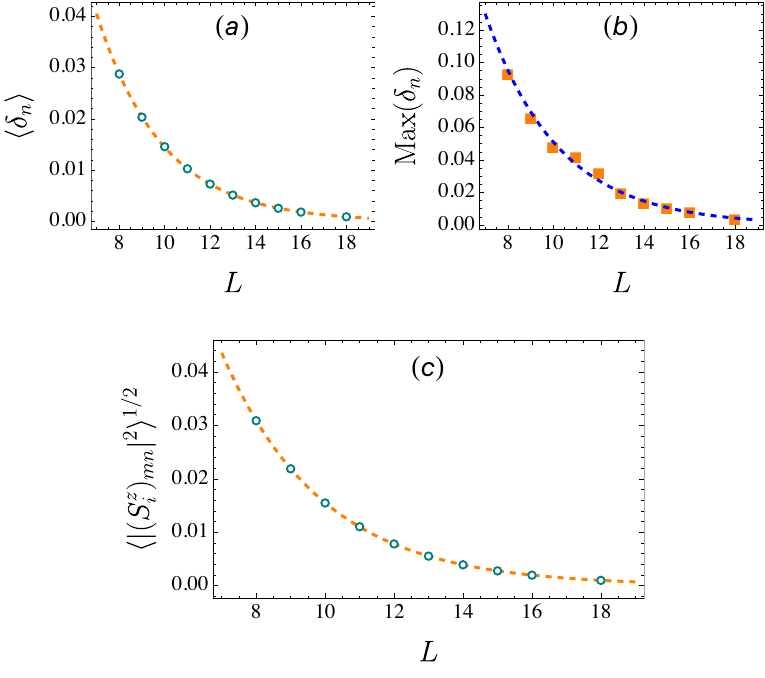}
    \caption{Scaling of fluctuation of diagonal matrix elements. $\delta_{n}= | (\sigma_{i}^{z})_{n+1}- (\sigma_{i}^{z})_{n} |$ where $i= L/2, (L+1)/2$ for $L$ even and odd respectively. (a) The scaling of average $\braket{\delta_{n}}$ with system size $L$. Off-diagonal elements $\delta_{mn}=  \braket{|(\sigma_{z}^{i})_{mn}|^{2}}^{1/2}$ , $i= \frac{L}{2},\frac{L+1}{2}$ for $L$ even and odd respectively.  (b) The scaling of maximum outliers, $\mathrm{Max}(\delta_{n})$ with system size $L$. (c) Scaling of average fluctuation of off-diagonal element $\braket{|(\sigma_{z}^{i})_{mn}|^{2}}^{1/2}$ , $i= \frac{L}{2},\frac{L+1}{2}$ for $L$ even and odd respectively. The dotted lines are the best fit line of the function $a_{0}2^{-L/b}$.}\label{fig:ethall}
\end{figure}

We study both the diagonal and the non-diagonal matrix elements. We choose our observable to be $\mathcal{O}= S^{i}_{z}$, i.e. $z$ component of the spin in the mid of the chain. $i$ is chosen to be $N/2, (N+1)/2$ when $N$ is even or odd respectively. To capture the variance of a diagonal element from its neighbor, we calculate the quantity \cite{PhysRevB.110.134206} : $\delta_{n}= | (\sigma_{z}^{i})_{n+1} - (\sigma_{z}^{i})_{n}|$ , $i= \frac{N}{2},\frac{N+1}{2}$. From the ETH ansatz this variation should scale as $\sim \rho(\bar{\phi})^{-1/2}$, and since $\rho (\bar{\phi}) \sim 2^{L}$ (this is justified for large systems), we thus expect $\braket{\delta_{n}} = a_{0} 2^{-L/2}$. We show the result for the same in Fig. \ref{fig:ethall} (a). The solid line is best fit to the expression $a_{0}2^{-L/b}$. We obtained the best fit parameters as $a_{0}= 0.447 \pm 0.003, b= 2.02 \pm 0.005$. Thus the ETH ansatz is followed closely by the KFIM. For finite-size systems however, it is possible that there are unusually large fluctuations around the mean-value, and the variance is thus high. We next examine the behavior of the maximum outlier of $\delta_{n}$; $\mathrm{Max} (\delta_{n})$. The behavior of maximum outliers are shown in Fig. \ref{fig:ethall} (b). The dashed line is the best fit line of the function  $a_{0}2^{-L/b}$. We obtained the best fit parameters as $a_{0}= 1.16 \pm 0.12, b= 2.22 \pm 0.06$. This suggests that the maximum outliers go decrease exponentially with system size and in the limit of infinite system size should go to zero, as is expected from the ETH ansatz.

For off-diagonal ETH we consider the average fluctuation of $\delta_{mn} \equiv  \braket{|(\sigma_{z}^{i})_{mn}|^{2}}^{1/2}$ , $i= \frac{L}{2},\frac{L+1}{2}$ for $L$ even and odd respectively. The result for which is shown in Fig. \ref{fig:ethall} (c) and the dashed line is the best line; $a_{0}2^{-L/b}$. We obtain the best fit parameters: $a_{0}= 0.486 \pm 0.001, b= 2.01 \pm 0.002$ which is close to the theoretical value of $b=2$ from the ETH ansatz. 

\begin{figure}[ht]
    \centering
    \includegraphics[width=  \linewidth]{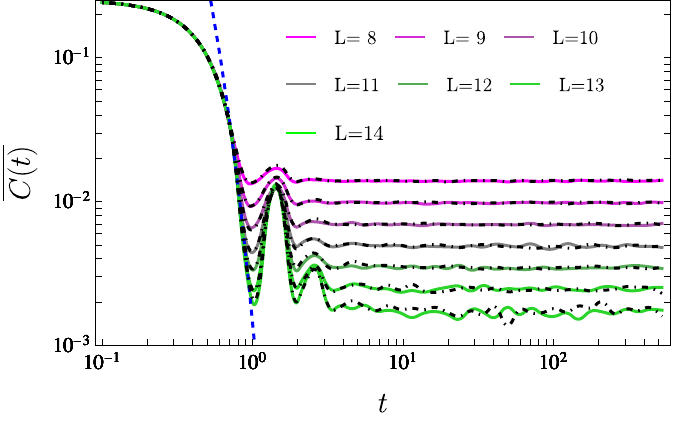}
    \caption{Average autocorrelation function given by Eq. \eqref{eq:autocor} for KFIM. Dashed black lines are the corresponding results for the COE random matrices. The dashed blue line is the best exponential fit for the decay part. }\label{fig:acfall}
\end{figure}

Finally, to study the dynamics of states we study the spin-spin autocorrelation defined as follows, for a single Hamiltonian realization:
\begin{equation}\label{eq:autocor}
    C(t)= \braket{\psi_{0}| \sigma^{z}_{L}(t)\sigma^{z}_{L}(0)|\psi_{0}},
\end{equation}
where the we take the $\ket{\psi_{0}}$ to be the eigenstate of $U_{\kfim}(U_{\mathrm{COE}})$. Finally, we take averaging over different Hamiltonian realization to obtain $\overline{C(t)}= \braket{C(t)}_{\{h_i\}}$, where $\braket{\cdot}_{\{h_i\}}$ here denotes ensemble averaging. The evolution of $\overline{C(t)}$ with time is shown in Fig. \ref{fig:acfall}. At early times it shows exponential decay. At late times it achieves a steady-state value that decreases with increasing system size $L$. Such behavior is a hallmark of ergodic system for which the correlations die at late times (going to zero in the large system size limit). The excellent agreement of the ETH ansatz, both for diagonal and off-diagonal ETH, and the characteristic decay and saturation of autocorrelation function suggest that the KFIM is indeed quantum chaotic.

\emph{Conclusion and outlook:} Using the prototypical example of the kicked field Ising model (KFIM), we show that, even though the system is quantum chaotic, there are features of random matrices that it fails to show. Notably we show that the maximum eigenvalue of the reduced density matrix for this do not follow the expected Tracy--Widom distribution. It is already shown in \cite{Herrmann:2024wat} that the distribution of the von Neumann entanglement entropy for  this system show deviation from expected COE random matrix ensemble. This study thus complements the study and shows that the reason for the deviation can be attributed to the largest eigenvalue of the density matrix that carries the largest weight when calculating the von Neumann entanglement entropy. We then show that instead of the Tracy--Widom distribution the largest eigenvalue follows the extreme value distribution. As was asserted in \cite{Herrmann:2024wat} that deviation in the von Neumann entanglement entropy distribution are due to the Hilbert space of this model which is a tensor product of $L$ two dimensional Hilbert space. We finally studied the thermalization properties of the KFIM. It is found that it follows the ETH ansatz quite well and also the autocorrelation function follows the expected ergodic behavior quite well. This thus suggests that the deviation as reported here and in \cite{Herrmann:2024wat} might not result in dramatic changes in the thermalization properties of the KFIM and other similar systems. The results also hints that the Gaussian and the circular matrix ensembles might be incapable to capture all the features of chaotic many-body quantum systems and that it might be necessary to introduce appropriate random matrix ensembles to properly model these many body systems.

\emph{Acknowledgments:}
We acknowledge Viktor Jahnke and Franco Nori for their comments.
Numerical computations were performed using the computational facilities of the Yukawa Institute for Theoretical Physics. TP acknowledges the partial support of the Yukawa Research Fellowship, supported by the Yukawa Memorial Foundation and JST CREST (Grant No. JPMJCR19T2).
The work was partially supported by JST CREST (Grant No. JPMJCR24I2).
M.~T. was partially supported by the Japan Society for the Promotion of Science (JSPS) Grants-in-Aid for Scientific Research (KAKENHI) Grants No. JP21H05185 and JP25K00925.
\bibliography{references}

\begin{thebibliography}{54}%
\makeatletter
\providecommand \@ifxundefined [1]{%
 \@ifx{#1\undefined}
}%
\providecommand \@ifnum [1]{%
 \ifnum #1\expandafter \@firstoftwo
 \else \expandafter \@secondoftwo
 \fi
}%
\providecommand \@ifx [1]{%
 \ifx #1\expandafter \@firstoftwo
 \else \expandafter \@secondoftwo
 \fi
}%
\providecommand \natexlab [1]{#1}%
\providecommand \enquote  [1]{``#1''}%
\providecommand \bibnamefont  [1]{#1}%
\providecommand \bibfnamefont [1]{#1}%
\providecommand \citenamefont [1]{#1}%
\providecommand \href@noop [0]{\@secondoftwo}%
\providecommand \href [0]{\begingroup \@sanitize@url \@href}%
\providecommand \@href[1]{\@@startlink{#1}\@@href}%
\providecommand \@@href[1]{\endgroup#1\@@endlink}%
\providecommand \@sanitize@url [0]{\catcode `\\12\catcode `\$12\catcode `\&12\catcode `\#12\catcode `\^12\catcode `\_12\catcode `\%12\relax}%
\providecommand \@@startlink[1]{}%
\providecommand \@@endlink[0]{}%
\providecommand \url  [0]{\begingroup\@sanitize@url \@url }%
\providecommand \@url [1]{\endgroup\@href {#1}{\urlprefix }}%
\providecommand \urlprefix  [0]{URL }%
\providecommand \Eprint [0]{\href }%
\providecommand \doibase [0]{http://dx.doi.org/}%
\providecommand \selectlanguage [0]{\@gobble}%
\providecommand \bibinfo  [0]{\@secondoftwo}%
\providecommand \bibfield  [0]{\@secondoftwo}%
\providecommand \translation [1]{[#1]}%
\providecommand \BibitemOpen [0]{}%
\providecommand \bibitemStop [0]{}%
\providecommand \bibitemNoStop [0]{.\EOS\space}%
\providecommand \EOS [0]{\spacefactor3000\relax}%
\providecommand \BibitemShut  [1]{\csname bibitem#1\endcsname}%
\let\auto@bib@innerbib\@empty
\bibitem [{\citenamefont {Bohigas}\ \emph {et~al.}(1984{\natexlab{a}})\citenamefont {Bohigas}, \citenamefont {Giannoni},\ and\ \citenamefont {Schmit}}]{PhysRevLett.52.1}%
  \BibitemOpen
  \bibfield  {author} {\bibinfo {author} {\bibfnamefont {O.}~\bibnamefont {Bohigas}}, \bibinfo {author} {\bibfnamefont {M.~J.}\ \bibnamefont {Giannoni}}, \ and\ \bibinfo {author} {\bibfnamefont {C.}~\bibnamefont {Schmit}},\ }\bibfield  {title} {\enquote {\bibinfo {title} {{Characterization of Chaotic Quantum Spectra and Universality of Level Fluctuation Laws}},}\ }\href {\doibase 10.1103/PhysRevLett.52.1} {\bibfield  {journal} {\bibinfo  {journal} {Phys. Rev. Lett.}\ }\textbf {\bibinfo {volume} {52}},\ \bibinfo {pages} {1--4} (\bibinfo {year} {1984}{\natexlab{a}})}\BibitemShut {NoStop}%
\bibitem [{\citenamefont {Bohigas}\ \emph {et~al.}(1984{\natexlab{b}})\citenamefont {Bohigas}, \citenamefont {Giannoni},\ and\ \citenamefont {Schmit}}]{bohigas1984spectral}%
  \BibitemOpen
  \bibfield  {author} {\bibinfo {author} {\bibfnamefont {O}~\bibnamefont {Bohigas}}, \bibinfo {author} {\bibfnamefont {MJ}~\bibnamefont {Giannoni}}, \ and\ \bibinfo {author} {\bibfnamefont {C}~\bibnamefont {Schmit}},\ }\bibfield  {title} {\enquote {\bibinfo {title} {Spectral properties of the laplacian and random matrix theories},}\ }\href {https://doi.org/10.1051/jphyslet:0198400450210101500} {\bibfield  {journal} {\bibinfo  {journal} {Journal de Physique Lettres}\ }\textbf {\bibinfo {volume} {45}},\ \bibinfo {pages} {1015--1022} (\bibinfo {year} {1984}{\natexlab{b}})}\BibitemShut {NoStop}%
\bibitem [{\citenamefont {Casati}\ \emph {et~al.}(1980)\citenamefont {Casati}, \citenamefont {Valz-Gris},\ and\ \citenamefont {Guarnieri}}]{casati1980connection}%
  \BibitemOpen
  \bibfield  {author} {\bibinfo {author} {\bibfnamefont {G}~\bibnamefont {Casati}}, \bibinfo {author} {\bibfnamefont {F}~\bibnamefont {Valz-Gris}}, \ and\ \bibinfo {author} {\bibfnamefont {I}~\bibnamefont {Guarnieri}},\ }\bibfield  {title} {\enquote {\bibinfo {title} {On the connection between quantization of nonintegrable systems and statistical theory of spectra},}\ }\href {https://doi.org/10.1007/BF02798790} {\bibfield  {journal} {\bibinfo  {journal} {Lettere al Nuovo Cimento (1971-1985)}\ }\textbf {\bibinfo {volume} {28}},\ \bibinfo {pages} {279--282} (\bibinfo {year} {1980})}\BibitemShut {NoStop}%
\bibitem [{\citenamefont {Berry}(1985)}]{berry1985semiclassical}%
  \BibitemOpen
  \bibfield  {author} {\bibinfo {author} {\bibfnamefont {Michael~Victor}\ \bibnamefont {Berry}},\ }\bibfield  {title} {\enquote {\bibinfo {title} {Semiclassical theory of spectral rigidity},}\ }\href {https://doi.org/10.1098/rspa.1985.0078} {\bibfield  {journal} {\bibinfo  {journal} {Proceedings of the Royal Society of London. A. Mathematical and Physical Sciences}\ }\textbf {\bibinfo {volume} {400}},\ \bibinfo {pages} {229--251} (\bibinfo {year} {1985})}\BibitemShut {NoStop}%
\bibitem [{\citenamefont {Mehta}(2004)}]{mehta2004random}%
  \BibitemOpen
  \bibfield  {author} {\bibinfo {author} {\bibfnamefont {M.~L.}\ \bibnamefont {Mehta}},\ }\href@noop {} {\emph {\bibinfo {title} {{Random Matrices}}}}\ (\bibinfo  {publisher} {Elsevier},\ \bibinfo {address} {San Diego, USA},\ \bibinfo {year} {2004})\BibitemShut {NoStop}%
\bibitem [{\citenamefont {Forrester}(2010)}]{forrester2010log}%
  \BibitemOpen
  \bibfield  {author} {\bibinfo {author} {\bibfnamefont {Peter~J}\ \bibnamefont {Forrester}},\ }\href@noop {} {\emph {\bibinfo {title} {Log-gases and random matrices (LMS-34)}}}\ (\bibinfo  {publisher} {Princeton university press},\ \bibinfo {year} {2010})\BibitemShut {NoStop}%
\bibitem [{\citenamefont {Berry}\ and\ \citenamefont {Tabor}(1977)}]{berry1977level}%
  \BibitemOpen
  \bibfield  {author} {\bibinfo {author} {\bibfnamefont {Michael~Victor}\ \bibnamefont {Berry}}\ and\ \bibinfo {author} {\bibfnamefont {Michael}\ \bibnamefont {Tabor}},\ }\bibfield  {title} {\enquote {\bibinfo {title} {Level clustering in the regular spectrum},}\ }\href {https://doi.org/10.1098/rspa.1977.0140} {\bibfield  {journal} {\bibinfo  {journal} {Proceedings of the Royal Society of London. A. Mathematical and Physical Sciences}\ }\textbf {\bibinfo {volume} {356}},\ \bibinfo {pages} {375--394} (\bibinfo {year} {1977})}\BibitemShut {NoStop}%
\bibitem [{\citenamefont {Marklof}(2001)}]{marklof2001berry}%
  \BibitemOpen
  \bibfield  {author} {\bibinfo {author} {\bibfnamefont {Jens}\ \bibnamefont {Marklof}},\ }\bibfield  {title} {\enquote {\bibinfo {title} {The {B}erry-{T}abor {c}onjecture},}\ }in\ \href {https://people.maths.bris.ac.uk/~majm/bib/3ecm.pdf} {\emph {\bibinfo {booktitle} {European Congress of Mathematics: Barcelona, July 10--14, 2000 Volume II}}}\ (\bibinfo {organization} {Springer},\ \bibinfo {year} {2001})\ pp.\ \bibinfo {pages} {421--427}\BibitemShut {NoStop}%
\bibitem [{\citenamefont {Hsu}\ and\ \citenamefont {Angle`s~d’Auriac}(1993)}]{Hsu_1993}%
  \BibitemOpen
  \bibfield  {author} {\bibinfo {author} {\bibfnamefont {Theodore~C.}\ \bibnamefont {Hsu}}\ and\ \bibinfo {author} {\bibfnamefont {J.~C.}\ \bibnamefont {Angle`s~d’Auriac}},\ }\bibfield  {title} {\enquote {\bibinfo {title} {Level repulsion in integrable and almost-integrable quantum spin models},}\ }\href {\doibase 10.1103/physrevb.47.14291} {\bibfield  {journal} {\bibinfo  {journal} {Phys. Rev. B}\ }\textbf {\bibinfo {volume} {47}},\ \bibinfo {pages} {14291–14296} (\bibinfo {year} {1993})},\ \Eprint {http://arxiv.org/abs/cond-mat/9211004} {arXiv:cond-mat/9211004 [cond-mat]} \BibitemShut {NoStop}%
\bibitem [{\citenamefont {Montambaux}\ \emph {et~al.}(1993)\citenamefont {Montambaux}, \citenamefont {Poilblanc}, \citenamefont {Bellissard},\ and\ \citenamefont {Sire}}]{PhysRevLett.70.497}%
  \BibitemOpen
  \bibfield  {author} {\bibinfo {author} {\bibfnamefont {Gilles}\ \bibnamefont {Montambaux}}, \bibinfo {author} {\bibfnamefont {Didier}\ \bibnamefont {Poilblanc}}, \bibinfo {author} {\bibfnamefont {Jean}\ \bibnamefont {Bellissard}}, \ and\ \bibinfo {author} {\bibfnamefont {Cl\'ement}\ \bibnamefont {Sire}},\ }\bibfield  {title} {\enquote {\bibinfo {title} {Quantum chaos in spin-fermion models},}\ }\href {\doibase 10.1103/PhysRevLett.70.497} {\bibfield  {journal} {\bibinfo  {journal} {Phys. Rev. Lett.}\ }\textbf {\bibinfo {volume} {70}},\ \bibinfo {pages} {497--500} (\bibinfo {year} {1993})}\BibitemShut {NoStop}%
\bibitem [{\citenamefont {Jacquod}\ and\ \citenamefont {Shepelyansky}(1997)}]{Jacquod_1997}%
  \BibitemOpen
  \bibfield  {author} {\bibinfo {author} {\bibfnamefont {Ph.}\ \bibnamefont {Jacquod}}\ and\ \bibinfo {author} {\bibfnamefont {D.~L.}\ \bibnamefont {Shepelyansky}},\ }\bibfield  {title} {\enquote {\bibinfo {title} {{Emergence of Quantum Chaos in Finite Interacting Fermi Systems}},}\ }\href {\doibase 10.1103/physrevlett.79.1837} {\bibfield  {journal} {\bibinfo  {journal} {Phys. Rev. Lett.}\ }\textbf {\bibinfo {volume} {79}},\ \bibinfo {pages} {1837–1840} (\bibinfo {year} {1997})},\ \Eprint {http://arxiv.org/abs/cond-mat/9706040} {arXiv:cond-mat/9706040 [cond-mat]} \BibitemShut {NoStop}%
\bibitem [{\citenamefont {Avishai}\ \emph {et~al.}(2002)\citenamefont {Avishai}, \citenamefont {Richert},\ and\ \citenamefont {Berkovits}}]{PhysRevB.66.052416}%
  \BibitemOpen
  \bibfield  {author} {\bibinfo {author} {\bibfnamefont {Y.}~\bibnamefont {Avishai}}, \bibinfo {author} {\bibfnamefont {J.}~\bibnamefont {Richert}}, \ and\ \bibinfo {author} {\bibfnamefont {R.}~\bibnamefont {Berkovits}},\ }\bibfield  {title} {\enquote {\bibinfo {title} {Level statistics in a heisenberg chain with random magnetic field},}\ }\href {\doibase 10.1103/PhysRevB.66.052416} {\bibfield  {journal} {\bibinfo  {journal} {Phys. Rev. B}\ }\textbf {\bibinfo {volume} {66}},\ \bibinfo {pages} {052416} (\bibinfo {year} {2002})}\BibitemShut {NoStop}%
\bibitem [{\citenamefont {Ba\~nuls}\ \emph {et~al.}(2011)\citenamefont {Ba\~nuls}, \citenamefont {Cirac},\ and\ \citenamefont {Hastings}}]{PhysRevLett.106.050405}%
  \BibitemOpen
  \bibfield  {author} {\bibinfo {author} {\bibfnamefont {M.~C.}\ \bibnamefont {Ba\~nuls}}, \bibinfo {author} {\bibfnamefont {J.~I.}\ \bibnamefont {Cirac}}, \ and\ \bibinfo {author} {\bibfnamefont {M.~B.}\ \bibnamefont {Hastings}},\ }\bibfield  {title} {\enquote {\bibinfo {title} {{Strong and Weak Thermalization of Infinite Nonintegrable Quantum Systems}},}\ }\href {\doibase 10.1103/PhysRevLett.106.050405} {\bibfield  {journal} {\bibinfo  {journal} {Phys. Rev. Lett.}\ }\textbf {\bibinfo {volume} {106}},\ \bibinfo {pages} {050405} (\bibinfo {year} {2011})}\BibitemShut {NoStop}%
\bibitem [{\citenamefont {Gharibyan}\ \emph {et~al.}(2018)\citenamefont {Gharibyan}, \citenamefont {Hanada}, \citenamefont {Shenker},\ and\ \citenamefont {Tezuka}}]{Gharibyan:2018jrp}%
  \BibitemOpen
  \bibfield  {author} {\bibinfo {author} {\bibfnamefont {Hrant}\ \bibnamefont {Gharibyan}}, \bibinfo {author} {\bibfnamefont {Masanori}\ \bibnamefont {Hanada}}, \bibinfo {author} {\bibfnamefont {Stephen~H.}\ \bibnamefont {Shenker}}, \ and\ \bibinfo {author} {\bibfnamefont {Masaki}\ \bibnamefont {Tezuka}},\ }\bibfield  {title} {\enquote {\bibinfo {title} {{Onset of Random Matrix Behavior in Scrambling Systems}},}\ }\href {\doibase 10.1007/JHEP07(2018)124} {\bibfield  {journal} {\bibinfo  {journal} {JHEP}\ }\textbf {\bibinfo {volume} {07}},\ \bibinfo {pages} {124} (\bibinfo {year} {2018})},\ \bibinfo {note} {[Erratum: JHEP 02, 197 (2019)]},\ \Eprint {http://arxiv.org/abs/1803.08050} {arXiv:1803.08050 [hep-th]} \BibitemShut {NoStop}%
\bibitem [{\citenamefont {Guhr}\ \emph {et~al.}(1998)\citenamefont {Guhr}, \citenamefont {Muller-Groeling},\ and\ \citenamefont {Weidenmuller}}]{Guhr:1997ve}%
  \BibitemOpen
  \bibfield  {author} {\bibinfo {author} {\bibfnamefont {Thomas}\ \bibnamefont {Guhr}}, \bibinfo {author} {\bibfnamefont {Axel}\ \bibnamefont {Muller-Groeling}}, \ and\ \bibinfo {author} {\bibfnamefont {Hans~A.}\ \bibnamefont {Weidenmuller}},\ }\bibfield  {title} {\enquote {\bibinfo {title} {{Random matrix theories in quantum physics: Common concepts}},}\ }\href {\doibase 10.1016/S0370-1573(97)00088-4} {\bibfield  {journal} {\bibinfo  {journal} {Phys. Rept.}\ }\textbf {\bibinfo {volume} {299}},\ \bibinfo {pages} {189--425} (\bibinfo {year} {1998})},\ \Eprint {http://arxiv.org/abs/cond-mat/9707301} {arXiv:cond-mat/9707301 [cond-mat]} \BibitemShut {NoStop}%
\bibitem [{\citenamefont {Eisert}\ \emph {et~al.}(2010)\citenamefont {Eisert}, \citenamefont {Cramer},\ and\ \citenamefont {Plenio}}]{RevModPhys.82.277}%
  \BibitemOpen
  \bibfield  {author} {\bibinfo {author} {\bibfnamefont {J.}~\bibnamefont {Eisert}}, \bibinfo {author} {\bibfnamefont {M.}~\bibnamefont {Cramer}}, \ and\ \bibinfo {author} {\bibfnamefont {M.~B.}\ \bibnamefont {Plenio}},\ }\bibfield  {title} {\enquote {\bibinfo {title} {Colloquium: Area laws for the entanglement entropy},}\ }\href {\doibase 10.1103/RevModPhys.82.277} {\bibfield  {journal} {\bibinfo  {journal} {Rev. Mod. Phys.}\ }\textbf {\bibinfo {volume} {82}},\ \bibinfo {pages} {277--306} (\bibinfo {year} {2010})}\BibitemShut {NoStop}%
\bibitem [{\citenamefont {Srdin\ifmmode~\check{s}\else \v{s}\fi{}ek}\ \emph {et~al.}(2021)\citenamefont {Srdin\ifmmode~\check{s}\else \v{s}\fi{}ek}, \citenamefont {Prosen},\ and\ \citenamefont {Sotiriadis}}]{PhysRevLett.126.121602}%
  \BibitemOpen
  \bibfield  {author} {\bibinfo {author} {\bibfnamefont {Miha}\ \bibnamefont {Srdin\ifmmode~\check{s}\else \v{s}\fi{}ek}}, \bibinfo {author} {\bibfnamefont {Toma{\ifmmode \check{z}\else \v{z}\fi{}}}\ \bibnamefont {Prosen}}, \ and\ \bibinfo {author} {\bibfnamefont {Spyros}\ \bibnamefont {Sotiriadis}},\ }\bibfield  {title} {\enquote {\bibinfo {title} {Signatures of chaos in nonintegrable models of quantum field theories},}\ }\href {\doibase 10.1103/PhysRevLett.126.121602} {\bibfield  {journal} {\bibinfo  {journal} {Phys. Rev. Lett.}\ }\textbf {\bibinfo {volume} {126}},\ \bibinfo {pages} {121602} (\bibinfo {year} {2021})},\ \Eprint {http://arxiv.org/abs/2012.08505} {arXiv:2012.08505 [cond-mat]} \BibitemShut {NoStop}%
\bibitem [{\citenamefont {B\"acker}\ \emph {et~al.}(2019)\citenamefont {B\"acker}, \citenamefont {Haque},\ and\ \citenamefont {Khaymovich}}]{PhysRevE.100.032117}%
  \BibitemOpen
  \bibfield  {author} {\bibinfo {author} {\bibfnamefont {Arnd}\ \bibnamefont {B\"acker}}, \bibinfo {author} {\bibfnamefont {Masudul}\ \bibnamefont {Haque}}, \ and\ \bibinfo {author} {\bibfnamefont {Ivan~M.}\ \bibnamefont {Khaymovich}},\ }\bibfield  {title} {\enquote {\bibinfo {title} {Multifractal dimensions for random matrices, chaotic quantum maps, and many-body systems},}\ }\href {\doibase 10.1103/PhysRevE.100.032117} {\bibfield  {journal} {\bibinfo  {journal} {Phys. Rev. E}\ }\textbf {\bibinfo {volume} {100}},\ \bibinfo {pages} {032117} (\bibinfo {year} {2019})},\ \Eprint {http://arxiv.org/abs/1905.03099} {arXiv:1905.03099 [cond-mat]} \BibitemShut {NoStop}%
\bibitem [{\citenamefont {Pausch}\ \emph {et~al.}(2021{\natexlab{a}})\citenamefont {Pausch}, \citenamefont {Carnio}, \citenamefont {Rodr\'{\i}guez},\ and\ \citenamefont {Buchleitner}}]{PhysRevLett.126.150601}%
  \BibitemOpen
  \bibfield  {author} {\bibinfo {author} {\bibfnamefont {Lukas}\ \bibnamefont {Pausch}}, \bibinfo {author} {\bibfnamefont {Edoardo~G.}\ \bibnamefont {Carnio}}, \bibinfo {author} {\bibfnamefont {Alberto}\ \bibnamefont {Rodr\'{\i}guez}}, \ and\ \bibinfo {author} {\bibfnamefont {Andreas}\ \bibnamefont {Buchleitner}},\ }\bibfield  {title} {\enquote {\bibinfo {title} {Chaos and ergodicity across the energy spectrum of interacting bosons},}\ }\href {\doibase 10.1103/PhysRevLett.126.150601} {\bibfield  {journal} {\bibinfo  {journal} {Phys. Rev. Lett.}\ }\textbf {\bibinfo {volume} {126}},\ \bibinfo {pages} {150601} (\bibinfo {year} {2021}{\natexlab{a}})},\ \Eprint {http://arxiv.org/abs/2009.05295} {arXiv:2009.05295 [cond-mat]} \BibitemShut {NoStop}%
\bibitem [{\citenamefont {Pausch}\ \emph {et~al.}(2021{\natexlab{b}})\citenamefont {Pausch}, \citenamefont {Carnio}, \citenamefont {Rodr\'\i{}guez},\ and\ \citenamefont {Buchleitner}}]{Pausch:2020pwc}%
  \BibitemOpen
  \bibfield  {author} {\bibinfo {author} {\bibfnamefont {Lukas}\ \bibnamefont {Pausch}}, \bibinfo {author} {\bibfnamefont {Edoardo~G.}\ \bibnamefont {Carnio}}, \bibinfo {author} {\bibfnamefont {Alberto}\ \bibnamefont {Rodr\'\i{}guez}}, \ and\ \bibinfo {author} {\bibfnamefont {Andreas}\ \bibnamefont {Buchleitner}},\ }\bibfield  {title} {\enquote {\bibinfo {title} {{Chaos and ergodicity across the energy spectrum of interacting bosons}},}\ }\href {\doibase 10.1103/PhysRevLett.126.150601} {\bibfield  {journal} {\bibinfo  {journal} {Phys. Rev. Lett.}\ }\textbf {\bibinfo {volume} {126}},\ \bibinfo {pages} {150601} (\bibinfo {year} {2021}{\natexlab{b}})},\ \Eprint {http://arxiv.org/abs/2009.05295} {arXiv:2009.05295 [cond-mat.quant-gas]} \BibitemShut {NoStop}%
\bibitem [{\citenamefont {Vidmar}\ \emph {et~al.}(2017)\citenamefont {Vidmar}, \citenamefont {Hackl}, \citenamefont {Bianchi},\ and\ \citenamefont {Rigol}}]{Vidmar:2017uux}%
  \BibitemOpen
  \bibfield  {author} {\bibinfo {author} {\bibfnamefont {Lev}\ \bibnamefont {Vidmar}}, \bibinfo {author} {\bibfnamefont {Lucas}\ \bibnamefont {Hackl}}, \bibinfo {author} {\bibfnamefont {Eugenio}\ \bibnamefont {Bianchi}}, \ and\ \bibinfo {author} {\bibfnamefont {Marcos}\ \bibnamefont {Rigol}},\ }\bibfield  {title} {\enquote {\bibinfo {title} {{Entanglement Entropy of Eigenstates of Quadratic Fermionic Hamiltonians}},}\ }\href {\doibase 10.1103/PhysRevLett.119.020601} {\bibfield  {journal} {\bibinfo  {journal} {Phys. Rev. Lett.}\ }\textbf {\bibinfo {volume} {119}},\ \bibinfo {pages} {020601} (\bibinfo {year} {2017})},\ \Eprint {http://arxiv.org/abs/1703.02979} {arXiv:1703.02979 [cond-mat.stat-mech]} \BibitemShut {NoStop}%
\bibitem [{\citenamefont {Vidmar}\ and\ \citenamefont {Rigol}(2017)}]{Vidmar:2017pak}%
  \BibitemOpen
  \bibfield  {author} {\bibinfo {author} {\bibfnamefont {Lev}\ \bibnamefont {Vidmar}}\ and\ \bibinfo {author} {\bibfnamefont {Marcos}\ \bibnamefont {Rigol}},\ }\bibfield  {title} {\enquote {\bibinfo {title} {{Entanglement Entropy of Eigenstates of Quantum Chaotic Hamiltonians}},}\ }\href {\doibase 10.1103/PhysRevLett.119.220603} {\bibfield  {journal} {\bibinfo  {journal} {Phys. Rev. Lett.}\ }\textbf {\bibinfo {volume} {119}},\ \bibinfo {pages} {220603} (\bibinfo {year} {2017})},\ \Eprint {http://arxiv.org/abs/1708.08453} {arXiv:1708.08453 [cond-mat.stat-mech]} \BibitemShut {NoStop}%
\bibitem [{\citenamefont {Haque}\ \emph {et~al.}(2022)\citenamefont {Haque}, \citenamefont {McClarty},\ and\ \citenamefont {Khaymovich}}]{Haque_2022}%
  \BibitemOpen
  \bibfield  {author} {\bibinfo {author} {\bibfnamefont {Masudul}\ \bibnamefont {Haque}}, \bibinfo {author} {\bibfnamefont {Paul~A.}\ \bibnamefont {McClarty}}, \ and\ \bibinfo {author} {\bibfnamefont {Ivan~M.}\ \bibnamefont {Khaymovich}},\ }\bibfield  {title} {\enquote {\bibinfo {title} {Entanglement of midspectrum eigenstates of chaotic many-body systems: Reasons for deviation from random ensembles},}\ }\href {http://dx.doi.org/10.1103/PhysRevE.105.014109} {\bibfield  {journal} {\bibinfo  {journal} {Phys. Rev. E}\ }\textbf {\bibinfo {volume} {105}} (\bibinfo {year} {2022})},\ \Eprint {http://arxiv.org/abs/2008.12782} {arXiv:2008.12782 [cond-mat]} \BibitemShut {NoStop}%
\bibitem [{\citenamefont {Huang}(2019)}]{Huang:2017std}%
  \BibitemOpen
  \bibfield  {author} {\bibinfo {author} {\bibfnamefont {Yichen}\ \bibnamefont {Huang}},\ }\bibfield  {title} {\enquote {\bibinfo {title} {{Universal eigenstate entanglement of chaotic local Hamiltonians}},}\ }\href {\doibase 10.1016/j.nuclphysb.2018.09.013} {\bibfield  {journal} {\bibinfo  {journal} {Nucl. Phys. B}\ }\textbf {\bibinfo {volume} {938}},\ \bibinfo {pages} {594--604} (\bibinfo {year} {2019})},\ \Eprint {http://arxiv.org/abs/1708.08607} {arXiv:1708.08607 [quant-ph]} \BibitemShut {NoStop}%
\bibitem [{\citenamefont {Huang}(2021)}]{huang2021universal}%
  \BibitemOpen
  \bibfield  {author} {\bibinfo {author} {\bibfnamefont {Yichen}\ \bibnamefont {Huang}},\ }\bibfield  {title} {\enquote {\bibinfo {title} {Universal entanglement of mid-spectrum eigenstates of chaotic local hamiltonians},}\ }\href {\doibase https://doi.org/10.1016/j.nuclphysb.2021.115373} {\bibfield  {journal} {\bibinfo  {journal} {Nuclear Physics B}\ }\textbf {\bibinfo {volume} {966}},\ \bibinfo {pages} {115373} (\bibinfo {year} {2021})}\BibitemShut {NoStop}%
\bibitem [{\citenamefont {Herrmann}\ \emph {et~al.}(2023)\citenamefont {Herrmann}, \citenamefont {Kieler},\ and\ \citenamefont {B\"acker}}]{Herrmann:2023hdj}%
  \BibitemOpen
  \bibfield  {author} {\bibinfo {author} {\bibfnamefont {Tabea}\ \bibnamefont {Herrmann}}, \bibinfo {author} {\bibfnamefont {Maximilian F.~I.}\ \bibnamefont {Kieler}}, \ and\ \bibinfo {author} {\bibfnamefont {Arnd}\ \bibnamefont {B\"acker}},\ }\bibfield  {title} {\enquote {\bibinfo {title} {{Characterizing quantum chaoticity of kicked spin chains}},}\ }\href {\doibase 10.1103/PhysRevE.108.044213} {\bibfield  {journal} {\bibinfo  {journal} {Phys. Rev. E}\ }\textbf {\bibinfo {volume} {108}},\ \bibinfo {pages} {044213} (\bibinfo {year} {2023})},\ \Eprint {http://arxiv.org/abs/2306.09034} {arXiv:2306.09034 [quant-ph]} \BibitemShut {NoStop}%
\bibitem [{\citenamefont {Bertini}\ \emph {et~al.}(2018)\citenamefont {Bertini}, \citenamefont {Kos},\ and\ \citenamefont {Prosen}}]{Bertini:2018wlu}%
  \BibitemOpen
  \bibfield  {author} {\bibinfo {author} {\bibfnamefont {Bruno}\ \bibnamefont {Bertini}}, \bibinfo {author} {\bibfnamefont {Pavel}\ \bibnamefont {Kos}}, \ and\ \bibinfo {author} {\bibfnamefont {Toma\v{z}}\ \bibnamefont {Prosen}},\ }\bibfield  {title} {\enquote {\bibinfo {title} {{Exact Spectral Form Factor in a Minimal Model of Many-Body Quantum Chaos}},}\ }\href {\doibase 10.1103/PhysRevLett.121.264101} {\bibfield  {journal} {\bibinfo  {journal} {Phys. Rev. Lett.}\ }\textbf {\bibinfo {volume} {121}},\ \bibinfo {pages} {264101} (\bibinfo {year} {2018})},\ \Eprint {http://arxiv.org/abs/1805.00931} {arXiv:1805.00931 [nlin.CD]} \BibitemShut {NoStop}%
\bibitem [{\citenamefont {Prosen}(2002)}]{PhysRevE.65.036208}%
  \BibitemOpen
  \bibfield  {author} {\bibinfo {author} {\bibfnamefont {Toma\v{z}}\ \bibnamefont {Prosen}},\ }\bibfield  {title} {\enquote {\bibinfo {title} {General relation between quantum ergodicity and fidelity of quantum dynamics},}\ }\href {\doibase 10.1103/PhysRevE.65.036208} {\bibfield  {journal} {\bibinfo  {journal} {Phys. Rev. E}\ }\textbf {\bibinfo {volume} {65}},\ \bibinfo {pages} {036208} (\bibinfo {year} {2002})},\ \Eprint {http://arxiv.org/abs/quant-ph/0106149} {arXiv:quant-ph/0106149 [quant-ph]} \BibitemShut {NoStop}%
\bibitem [{\citenamefont {Bertini}\ \emph {et~al.}(2019)\citenamefont {Bertini}, \citenamefont {Kos},\ and\ \citenamefont {Prosen}}]{Bertini:2018fbz}%
  \BibitemOpen
  \bibfield  {author} {\bibinfo {author} {\bibfnamefont {Bruno}\ \bibnamefont {Bertini}}, \bibinfo {author} {\bibfnamefont {Pavel}\ \bibnamefont {Kos}}, \ and\ \bibinfo {author} {\bibfnamefont {Toma\v{z}}\ \bibnamefont {Prosen}},\ }\bibfield  {title} {\enquote {\bibinfo {title} {{Entanglement spreading in a minimal model of maximal many-body quantum chaos}},}\ }\href {\doibase 10.1103/PhysRevX.9.021033} {\bibfield  {journal} {\bibinfo  {journal} {Phys. Rev. X}\ }\textbf {\bibinfo {volume} {9}},\ \bibinfo {pages} {021033} (\bibinfo {year} {2019})},\ \Eprint {http://arxiv.org/abs/1812.05090} {arXiv:1812.05090 [cond-mat.stat-mech]} \BibitemShut {NoStop}%
\bibitem [{\citenamefont {Braun}\ \emph {et~al.}(2020)\citenamefont {Braun}, \citenamefont {Waltner}, \citenamefont {Akila}, \citenamefont {Gutkin},\ and\ \citenamefont {Guhr}}]{Braun:2019kgf}%
  \BibitemOpen
  \bibfield  {author} {\bibinfo {author} {\bibfnamefont {Petr}\ \bibnamefont {Braun}}, \bibinfo {author} {\bibfnamefont {Daniel}\ \bibnamefont {Waltner}}, \bibinfo {author} {\bibfnamefont {Maram}\ \bibnamefont {Akila}}, \bibinfo {author} {\bibfnamefont {Boris}\ \bibnamefont {Gutkin}}, \ and\ \bibinfo {author} {\bibfnamefont {Thomas}\ \bibnamefont {Guhr}},\ }\bibfield  {title} {\enquote {\bibinfo {title} {{Transition from quantum chaos to localization in spin chains}},}\ }\href {\doibase 10.1103/PhysRevE.101.052201} {\bibfield  {journal} {\bibinfo  {journal} {Phys. Rev. E}\ }\textbf {\bibinfo {volume} {101}},\ \bibinfo {pages} {052201} (\bibinfo {year} {2020})},\ \Eprint {http://arxiv.org/abs/1902.06265} {arXiv:1902.06265 [cond-mat.stat-mech]} \BibitemShut {NoStop}%
\bibitem [{\citenamefont {Flack}\ \emph {et~al.}(2020)\citenamefont {Flack}, \citenamefont {Bertini},\ and\ \citenamefont {Prosen}}]{Flack:2020ybm}%
  \BibitemOpen
  \bibfield  {author} {\bibinfo {author} {\bibfnamefont {Ana}\ \bibnamefont {Flack}}, \bibinfo {author} {\bibfnamefont {Bruno}\ \bibnamefont {Bertini}}, \ and\ \bibinfo {author} {\bibfnamefont {Toma\v{z}}\ \bibnamefont {Prosen}},\ }\bibfield  {title} {\enquote {\bibinfo {title} {{Statistics of the spectral form factor in the self-dual kicked Ising model}},}\ }\href {\doibase 10.1103/PhysRevResearch.2.043403} {\bibfield  {journal} {\bibinfo  {journal} {Phys. Rev. Res.}\ }\textbf {\bibinfo {volume} {2}},\ \bibinfo {pages} {043403} (\bibinfo {year} {2020})},\ \Eprint {http://arxiv.org/abs/2009.03199} {arXiv:2009.03199 [nlin.CD]} \BibitemShut {NoStop}%
\bibitem [{\citenamefont {Herrmann}\ \emph {et~al.}(2025)\citenamefont {Herrmann}, \citenamefont {Brandau},\ and\ \citenamefont {B\"acker}}]{Herrmann:2024wat}%
  \BibitemOpen
  \bibfield  {author} {\bibinfo {author} {\bibfnamefont {Tabea}\ \bibnamefont {Herrmann}}, \bibinfo {author} {\bibfnamefont {Roland}\ \bibnamefont {Brandau}}, \ and\ \bibinfo {author} {\bibfnamefont {Arnd}\ \bibnamefont {B\"acker}},\ }\bibfield  {title} {\enquote {\bibinfo {title} {{Deviations from random-matrix entanglement statistics for kicked quantum chaotic spin-1/2 chains}},}\ }\href {\doibase 10.1103/PhysRevE.111.L012104} {\bibfield  {journal} {\bibinfo  {journal} {Phys. Rev. E}\ }\textbf {\bibinfo {volume} {111}},\ \bibinfo {pages} {L012104} (\bibinfo {year} {2025})},\ \Eprint {http://arxiv.org/abs/2405.07545} {arXiv:2405.07545 [quant-ph]} \BibitemShut {NoStop}%
\bibitem [{\citenamefont {Akila}\ \emph {et~al.}(2016)\citenamefont {Akila}, \citenamefont {Waltner}, \citenamefont {Gutkin},\ and\ \citenamefont {Guhr}}]{akila2016particle}%
  \BibitemOpen
  \bibfield  {author} {\bibinfo {author} {\bibfnamefont {Maram}\ \bibnamefont {Akila}}, \bibinfo {author} {\bibfnamefont {Daniel}\ \bibnamefont {Waltner}}, \bibinfo {author} {\bibfnamefont {Boris}\ \bibnamefont {Gutkin}}, \ and\ \bibinfo {author} {\bibfnamefont {Thomas}\ \bibnamefont {Guhr}},\ }\bibfield  {title} {\enquote {\bibinfo {title} {Particle-time duality in the kicked ising spin chain},}\ }\href {\doibase https://doi.org/10.1088/1751-8113/49/37/375101} {\bibfield  {journal} {\bibinfo  {journal} {Journal of Physics A: Mathematical and Theoretical}\ }\textbf {\bibinfo {volume} {49}},\ \bibinfo {pages} {375101} (\bibinfo {year} {2016})},\ \Eprint {http://arxiv.org/abs/1602.07130} {arXiv:1602.07130 [nlin]} \BibitemShut {NoStop}%
\bibitem [{\citenamefont {Kos}\ \emph {et~al.}(2018)\citenamefont {Kos}, \citenamefont {Ljubotina},\ and\ \citenamefont {Prosen}}]{Kos:2017zjh}%
  \BibitemOpen
  \bibfield  {author} {\bibinfo {author} {\bibfnamefont {Pavel}\ \bibnamefont {Kos}}, \bibinfo {author} {\bibfnamefont {Marko}\ \bibnamefont {Ljubotina}}, \ and\ \bibinfo {author} {\bibfnamefont {Tomaz}\ \bibnamefont {Prosen}},\ }\bibfield  {title} {\enquote {\bibinfo {title} {Many-body quantum chaos: Analytic connection to random matrix theory},}\ }\href {\doibase 10.1103/PhysRevX.8.021062} {\bibfield  {journal} {\bibinfo  {journal} {Phys. Rev. X}\ }\textbf {\bibinfo {volume} {8}},\ \bibinfo {pages} {021062} (\bibinfo {year} {2018})},\ \Eprint {http://arxiv.org/abs/1712.02665} {arXiv:1712.02665 [nlin.CD]} \BibitemShut {NoStop}%
\bibitem [{\citenamefont {Lerose}\ \emph {et~al.}(2021)\citenamefont {Lerose}, \citenamefont {Sonner},\ and\ \citenamefont {Abanin}}]{Lerose:2020fhd}%
  \BibitemOpen
  \bibfield  {author} {\bibinfo {author} {\bibfnamefont {Alessio}\ \bibnamefont {Lerose}}, \bibinfo {author} {\bibfnamefont {Michael}\ \bibnamefont {Sonner}}, \ and\ \bibinfo {author} {\bibfnamefont {Dmitry~A.}\ \bibnamefont {Abanin}},\ }\bibfield  {title} {\enquote {\bibinfo {title} {{Influence Matrix Approach to Many-Body Floquet Dynamics}},}\ }\href {\doibase 10.1103/PhysRevX.11.021040} {\bibfield  {journal} {\bibinfo  {journal} {Phys. Rev. X}\ }\textbf {\bibinfo {volume} {11}},\ \bibinfo {pages} {021040} (\bibinfo {year} {2021})},\ \Eprint {http://arxiv.org/abs/2009.10105} {arXiv:2009.10105 [cond-mat.str-el]} \BibitemShut {NoStop}%
\bibitem [{\citenamefont {Duenez}(2004)}]{duenez2004random}%
  \BibitemOpen
  \bibfield  {author} {\bibinfo {author} {\bibfnamefont {Eduardo}\ \bibnamefont {Duenez}},\ }\bibfield  {title} {\enquote {\bibinfo {title} {Random matrix ensembles associated to compact symmetric spaces},}\ }\href@noop {} {\bibfield  {journal} {\bibinfo  {journal} {Communications in mathematical physics}\ }\textbf {\bibinfo {volume} {244}},\ \bibinfo {pages} {29--61} (\bibinfo {year} {2004})},\ \Eprint {http://arxiv.org/abs/math-ph/0111005} {arXiv:math-ph/0111005 [math-ph]} \BibitemShut {NoStop}%
\bibitem [{\citenamefont {Porter}(1965)}]{porter1965statistical}%
  \BibitemOpen
  \bibfield  {author} {\bibinfo {author} {\bibfnamefont {Charles~E}\ \bibnamefont {Porter}},\ }\bibfield  {title} {\enquote {\bibinfo {title} {{Statistical theories of spectra: Fluctuations}},}\ }\href@noop {} {\bibfield  {journal} {\bibinfo  {journal} {(Academic, New York).}\ } (\bibinfo {year} {1965})}\BibitemShut {NoStop}%
\bibitem [{\citenamefont {Laflorencie}(2016)}]{Laflorencie_2016}%
  \BibitemOpen
  \bibfield  {author} {\bibinfo {author} {\bibfnamefont {Nicolas}\ \bibnamefont {Laflorencie}},\ }\bibfield  {title} {\enquote {\bibinfo {title} {Quantum entanglement in condensed matter systems},}\ }\href {\doibase 10.1016/j.physrep.2016.06.008} {\bibfield  {journal} {\bibinfo  {journal} {Physics Reports}\ }\textbf {\bibinfo {volume} {646}},\ \bibinfo {pages} {1–59} (\bibinfo {year} {2016})},\ \Eprint {http://arxiv.org/abs/1512.03388} {arXiv:1512.03388 [cond-mat]} \BibitemShut {NoStop}%
\bibitem [{\citenamefont {Luitz}(2021)}]{Luitz_2021}%
  \BibitemOpen
  \bibfield  {author} {\bibinfo {author} {\bibfnamefont {David~J.}\ \bibnamefont {Luitz}},\ }\bibfield  {title} {\enquote {\bibinfo {title} {Polynomial filter diagonalization of large floquet unitary operators},}\ }\href {http://dx.doi.org/10.21468/SciPostPhys.11.2.021} {\bibfield  {journal} {\bibinfo  {journal} {SciPost Physics}\ }\textbf {\bibinfo {volume} {11}} (\bibinfo {year} {2021})},\ \Eprint {http://arxiv.org/abs/2102.05054} {arXiv:2102.05054 [cond-mat]} \BibitemShut {NoStop}%
\bibitem [{\citenamefont {Sierant}\ \emph {et~al.}(2023)\citenamefont {Sierant}, \citenamefont {Lewenstein}, \citenamefont {Scardicchio},\ and\ \citenamefont {Zakrzewski}}]{Sierant:2022xtl}%
  \BibitemOpen
  \bibfield  {author} {\bibinfo {author} {\bibfnamefont {Piotr}\ \bibnamefont {Sierant}}, \bibinfo {author} {\bibfnamefont {Maciej}\ \bibnamefont {Lewenstein}}, \bibinfo {author} {\bibfnamefont {Antonello}\ \bibnamefont {Scardicchio}}, \ and\ \bibinfo {author} {\bibfnamefont {Jakub}\ \bibnamefont {Zakrzewski}},\ }\bibfield  {title} {\enquote {\bibinfo {title} {{Stability of many-body localization in Floquet systems}},}\ }\href {\doibase 10.1103/PhysRevB.107.115132} {\bibfield  {journal} {\bibinfo  {journal} {Phys. Rev. B}\ }\textbf {\bibinfo {volume} {107}},\ \bibinfo {pages} {115132} (\bibinfo {year} {2023})},\ \Eprint {http://arxiv.org/abs/2203.15697} {arXiv:2203.15697 [cond-mat.dis-nn]} \BibitemShut {NoStop}%
\bibitem [{sup()}]{supp}%
  \BibitemOpen
  \href@noop {} {}\bibinfo {note} {See Supplemental Material at URL-will-be-inserted-by-publisher, with additional references \cite{dumitriu2002matrix,larkin1969quasiclassical,Maldacena:2015waa,Hashimoto:2017oit,matrixfreemma}, for additional supporting results.}\BibitemShut {Stop}%
\bibitem [{\citenamefont {Tao}\ and\ \citenamefont {Vu}(2012)}]{tao2012random}%
  \BibitemOpen
  \bibfield  {author} {\bibinfo {author} {\bibfnamefont {Terence}\ \bibnamefont {Tao}}\ and\ \bibinfo {author} {\bibfnamefont {Van}\ \bibnamefont {Vu}},\ }\bibfield  {title} {\enquote {\bibinfo {title} {{Random covariance matrices: Universality of local statistics of eigenvalues}},}\ }\href {\doibase 10.1214/11-AOP648} {\bibfield  {journal} {\bibinfo  {journal} {The Annals of Probability}\ }\textbf {\bibinfo {volume} {40}},\ \bibinfo {pages} {1285--1315} (\bibinfo {year} {2012})},\ \Eprint {http://arxiv.org/abs/0912.0966} {arXiv:0912.0966 [math]} \BibitemShut {NoStop}%
\bibitem [{\citenamefont {Johnstone}(2001)}]{johnstone2001}%
  \BibitemOpen
  \bibfield  {author} {\bibinfo {author} {\bibfnamefont {Iain~M.}\ \bibnamefont {Johnstone}},\ }\bibfield  {title} {\enquote {\bibinfo {title} {{On the distribution of the largest eigenvalue in principal components analysis}},}\ }\href {\doibase 10.1214/aos/1009210544} {\bibfield  {journal} {\bibinfo  {journal} {The Annals of Statistics}\ }\textbf {\bibinfo {volume} {29}},\ \bibinfo {pages} {295--327} (\bibinfo {year} {2001})}\BibitemShut {NoStop}%
\bibitem [{\citenamefont {Fisher}\ and\ \citenamefont {Tippett}(1928)}]{fisher1928limiting}%
  \BibitemOpen
  \bibfield  {author} {\bibinfo {author} {\bibfnamefont {Ronald~Aylmer}\ \bibnamefont {Fisher}}\ and\ \bibinfo {author} {\bibfnamefont {Leonard Henry~Caleb}\ \bibnamefont {Tippett}},\ }\bibfield  {title} {\enquote {\bibinfo {title} {{Limiting forms of the frequency distribution of the largest or smallest member of a sample}},}\ }in\ \href@noop {} {\emph {\bibinfo {booktitle} {Mathematical proceedings of the Cambridge philosophical society}}},\ Vol.~\bibinfo {volume} {24}\ (\bibinfo {organization} {Cambridge University Press},\ \bibinfo {year} {1928})\ pp.\ \bibinfo {pages} {180--190}\BibitemShut {NoStop}%
\bibitem [{\citenamefont {Gumbel}(1958)}]{gumbel1958}%
  \BibitemOpen
  \bibfield  {author} {\bibinfo {author} {\bibfnamefont {Emil~Julius}\ \bibnamefont {Gumbel}},\ }\href@noop {} {\emph {\bibinfo {title} {{Statistics of Extremes}}}}\ (\bibinfo  {publisher} {Columbia University Press},\ \bibinfo {address} {New York},\ \bibinfo {year} {1958})\BibitemShut {NoStop}%
\bibitem [{\citenamefont {Majumdar}\ \emph {et~al.}(2020)\citenamefont {Majumdar}, \citenamefont {Pal},\ and\ \citenamefont {Schehr}}]{Majumdar_2020}%
  \BibitemOpen
  \bibfield  {author} {\bibinfo {author} {\bibfnamefont {Satya~N.}\ \bibnamefont {Majumdar}}, \bibinfo {author} {\bibfnamefont {Arnab}\ \bibnamefont {Pal}}, \ and\ \bibinfo {author} {\bibfnamefont {Grégory}\ \bibnamefont {Schehr}},\ }\bibfield  {title} {\enquote {\bibinfo {title} {Extreme value statistics of correlated random variables: A pedagogical review},}\ }\href {\doibase 10.1016/j.physrep.2019.10.005} {\bibfield  {journal} {\bibinfo  {journal} {Physics Reports}\ }\textbf {\bibinfo {volume} {840}},\ \bibinfo {pages} {1–32} (\bibinfo {year} {2020})},\ \Eprint {http://arxiv.org/abs/1910.10667} {arXiv:1910.10667 [cond-mat]} \BibitemShut {NoStop}%
\bibitem [{\citenamefont {Leadbetter}\ and\ \citenamefont {Rootzen}(1988)}]{leadbetter1988extremal}%
  \BibitemOpen
  \bibfield  {author} {\bibinfo {author} {\bibfnamefont {M.~R.}\ \bibnamefont {Leadbetter}}\ and\ \bibinfo {author} {\bibfnamefont {Holger}\ \bibnamefont {Rootzen}},\ }\bibfield  {title} {\enquote {\bibinfo {title} {{Extremal Theory for Stochastic Processes}},}\ }\href {\doibase 10.1214/aop/1176991767} {\bibfield  {journal} {\bibinfo  {journal} {The Annals of Probability}\ }\textbf {\bibinfo {volume} {16}},\ \bibinfo {pages} {431--478} (\bibinfo {year} {1988})}\BibitemShut {NoStop}%
\bibitem [{\citenamefont {Rodriguez-Nieva}\ \emph {et~al.}(2023)\citenamefont {Rodriguez-Nieva}, \citenamefont {Jonay},\ and\ \citenamefont {Khemani}}]{Rodriguez-Nieva:2023err}%
  \BibitemOpen
  \bibfield  {author} {\bibinfo {author} {\bibfnamefont {Joaquin~F.}\ \bibnamefont {Rodriguez-Nieva}}, \bibinfo {author} {\bibfnamefont {Cheryne}\ \bibnamefont {Jonay}}, \ and\ \bibinfo {author} {\bibfnamefont {Vedika}\ \bibnamefont {Khemani}},\ }\bibfield  {title} {\enquote {\bibinfo {title} {{Quantifying quantum chaos through microcanonical distributions of entanglement}},}\ }\href@noop {} {\  (\bibinfo {year} {2023})},\ \Eprint {http://arxiv.org/abs/2305.11940} {arXiv:2305.11940 [cond-mat.stat-mech]} \BibitemShut {NoStop}%
\bibitem [{\citenamefont {Kliczkowski}\ \emph {et~al.}(2024)\citenamefont {Kliczkowski}, \citenamefont {\ifmmode~\acute{S}\else \'Swiętek}, \citenamefont {Hopjan},\ and\ \citenamefont {Vidmar}}]{PhysRevB.110.134206}%
  \BibitemOpen
  \bibfield  {author} {\bibinfo {author} {\bibfnamefont {Maksymilian}\ \bibnamefont {Kliczkowski}}, \bibinfo {author} {\bibfnamefont {Rafa\l{}}\ \bibnamefont {\ifmmode~\acute{S}\else \'Swiętek}}, \bibinfo {author} {\bibfnamefont {Miroslav}\ \bibnamefont {Hopjan}}, \ and\ \bibinfo {author} {\bibfnamefont {Lev}\ \bibnamefont {Vidmar}},\ }\bibfield  {title} {\enquote {\bibinfo {title} {{Fading ergodicity}},}\ }\href {\doibase 10.1103/PhysRevB.110.134206} {\bibfield  {journal} {\bibinfo  {journal} {Phys. Rev. B}\ }\textbf {\bibinfo {volume} {110}},\ \bibinfo {pages} {134206} (\bibinfo {year} {2024})},\ \Eprint {http://arxiv.org/abs/2407.16773} {arXiv:2407.16773 [cond-mat]} \BibitemShut {NoStop}%
\bibitem [{\citenamefont {Dumitriu}\ and\ \citenamefont {Edelman}(2002)}]{dumitriu2002matrix}%
  \BibitemOpen
  \bibfield  {author} {\bibinfo {author} {\bibfnamefont {Ioana}\ \bibnamefont {Dumitriu}}\ and\ \bibinfo {author} {\bibfnamefont {Alan}\ \bibnamefont {Edelman}},\ }\bibfield  {title} {\enquote {\bibinfo {title} {{Matrix models for beta ensembles}},}\ }\href {https://doi.org/10.1063/1.1507823} {\bibfield  {journal} {\bibinfo  {journal} {Journal of Mathematical Physics}\ }\textbf {\bibinfo {volume} {43}},\ \bibinfo {pages} {5830--5847} (\bibinfo {year} {2002})}\BibitemShut {NoStop}%
\bibitem [{\citenamefont {Larkin}\ and\ \citenamefont {Ovchinnikov}(1969)}]{larkin1969quasiclassical}%
  \BibitemOpen
  \bibfield  {author} {\bibinfo {author} {\bibfnamefont {Anatoly~I}\ \bibnamefont {Larkin}}\ and\ \bibinfo {author} {\bibfnamefont {Yu~N}\ \bibnamefont {Ovchinnikov}},\ }\bibfield  {title} {\enquote {\bibinfo {title} {Quasiclassical method in the theory of superconductivity},}\ }\href@noop {} {\bibfield  {journal} {\bibinfo  {journal} {Sov Phys JETP}\ }\textbf {\bibinfo {volume} {28}},\ \bibinfo {pages} {1200--1205} (\bibinfo {year} {1969})}\BibitemShut {NoStop}%
\bibitem [{\citenamefont {Maldacena}\ \emph {et~al.}(2016)\citenamefont {Maldacena}, \citenamefont {Shenker},\ and\ \citenamefont {Stanford}}]{Maldacena:2015waa}%
  \BibitemOpen
  \bibfield  {author} {\bibinfo {author} {\bibfnamefont {Juan}\ \bibnamefont {Maldacena}}, \bibinfo {author} {\bibfnamefont {Stephen~H.}\ \bibnamefont {Shenker}}, \ and\ \bibinfo {author} {\bibfnamefont {Douglas}\ \bibnamefont {Stanford}},\ }\bibfield  {title} {\enquote {\bibinfo {title} {{A bound on chaos}},}\ }\href {\doibase 10.1007/JHEP08(2016)106} {\bibfield  {journal} {\bibinfo  {journal} {JHEP}\ }\textbf {\bibinfo {volume} {08}},\ \bibinfo {pages} {106} (\bibinfo {year} {2016})},\ \Eprint {http://arxiv.org/abs/1503.01409} {arXiv:1503.01409 [hep-th]} \BibitemShut {NoStop}%
\bibitem [{\citenamefont {Hashimoto}\ \emph {et~al.}(2017)\citenamefont {Hashimoto}, \citenamefont {Murata},\ and\ \citenamefont {Yoshii}}]{Hashimoto:2017oit}%
  \BibitemOpen
  \bibfield  {author} {\bibinfo {author} {\bibfnamefont {Koji}\ \bibnamefont {Hashimoto}}, \bibinfo {author} {\bibfnamefont {Keiju}\ \bibnamefont {Murata}}, \ and\ \bibinfo {author} {\bibfnamefont {Ryosuke}\ \bibnamefont {Yoshii}},\ }\bibfield  {title} {\enquote {\bibinfo {title} {{Out-of-time-order correlators in quantum mechanics}},}\ }\href {\doibase 10.1007/JHEP10(2017)138} {\bibfield  {journal} {\bibinfo  {journal} {JHEP}\ }\textbf {\bibinfo {volume} {10}},\ \bibinfo {pages} {138} (\bibinfo {year} {2017})},\ \Eprint {http://arxiv.org/abs/1703.09435} {arXiv:1703.09435 [hep-th]} \BibitemShut {NoStop}%
\bibitem [{\citenamefont {Schumacher}(2025)}]{matrixfreemma}%
  \BibitemOpen
  \bibfield  {author} {\bibinfo {author} {\bibfnamefont {Henrik}\ \bibnamefont {Schumacher}},\ }\href {https://mathematica.stackexchange.com/a/311316/73763} {\enquote {\bibinfo {title} {Efficient way to do(or matrix free) arnoldi},}\ } (\bibinfo {year} {2025}),\ \bibinfo {note} {answer on Mathematica Stack Exchange}\BibitemShut {NoStop}%
\end{thebibliography}%


\newpage

\hbox{}\thispagestyle{empty}\newpage
\appendix\label{appendix}
\onecolumngrid
\begin{center}
\textbf{\large Supplemental Materials: \\Extreme value statistics and eigenstate thermalization in kicked quantum chaotic spin-$1/2$ chains}
\vspace{2ex}

Tanay Pathak and Masaki Tezuka
\end{center}

\section{Marchenko--Pastur distribution}\label{sec:mplaw}
\begin{figure}[ht]
    \centering
    \includegraphics[width=  0.8\linewidth]{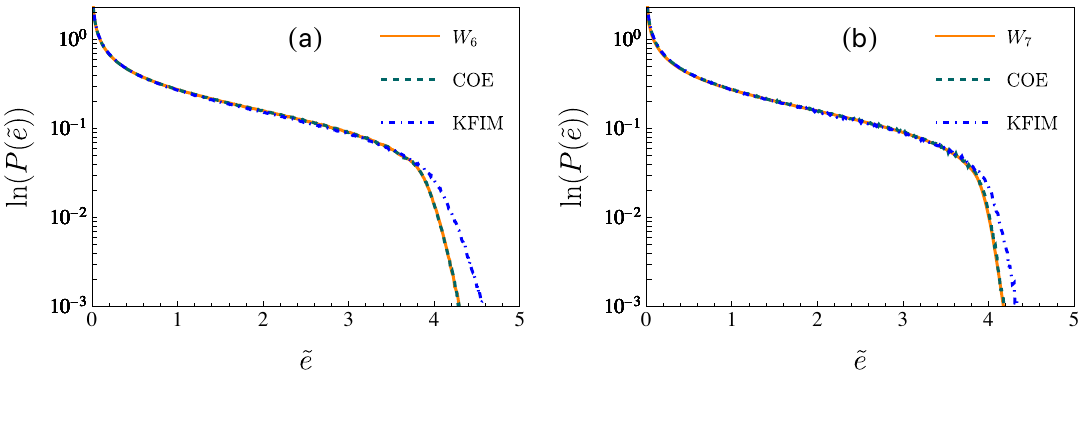}
    \caption{
    (a) Marchenko--Pastur distribution for Wishart matrices of size $2^{6} \times 2^{6}$, COE matrix and KFIM for $N =12$. (b) Marchenko--Pastur distribution for Wishart matrices of size $2^{7} \times 2^{7}$, COE matrix and KFIM for $N =14$. Observe the deviation towards the tail for the KFIM in both the figures, while the corresponding result of the COE matrix agrees well with the result for Wishart matrices. These deviations become less prominent as we increase $L$.}\label{fig:mpcompare}
\end{figure}

In this section we show some extra result of the comparison of Marchenko--Pastur distribution, obtained using finite size Wishart matrices, with the numerical distribution of Schmidt values obtained for KFIM and COE random matrices. It is already shown in the main section that there are differences between the result of KFIM and the Marchenko--Pastur distribution for finite dimensional matrices. In Fig. \ref{fig:mpcompare} we however show that for a COE random matrix of same dimension the results of Wishart matrices agree. We show the results for $L = 12 (300000), 14 (101600)$, for both the COE matrix and the KFIM, where the number in the brackets denotes the total number of eigenstates considered. We will see in the following section that this agreement between the result of the Wishart matrices and COE matrices are seen in the distribution of largest eigenvalue as well. 

\section{Tracy--Widom distribution}\label{sec:tw}

\begin{figure}[ht]
    \centering
    \includegraphics[width=  0.8\linewidth]{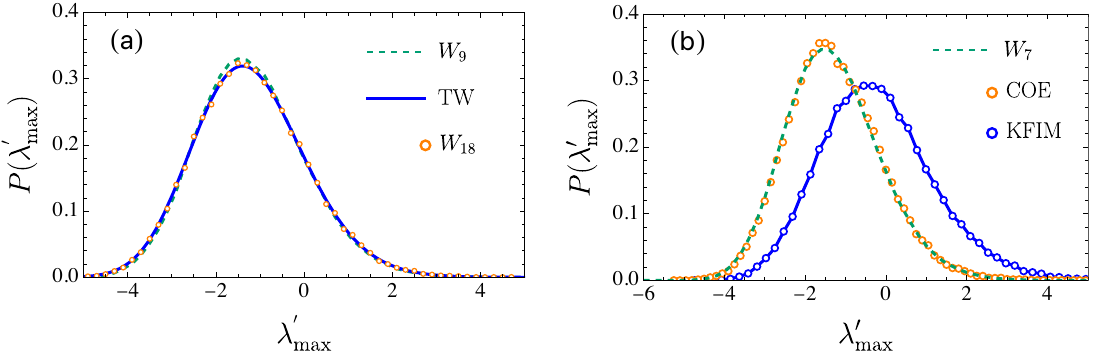}
    \caption{(a) Comparison of numerically obtained distribution for real Wishart matrices with the standard Tracy--Widom distribution of order 1. We take the Wishart matrices of dimensions= $2^{9} \times 2^{9}(10^6)$ and $2^{18} \times 2^{18}$ (50000), numbers in bracket denote number of realizations considered. (b) Comparison of numerically obtained distribution of $\lambda'_{\max}$ with COE and KFIM, $N= 14$. Half of the system is traced out, with remaining subsystem dimension $= 2^{7}$. Observe the nice agreement between COE and Wishart matrices result and clear deviation of the KFIM for the same.}\label{fig:twcom}
\end{figure}

Tracy--Widom distribution is only valid asymptotically in the limit of large matrix dimensions \cite{johnstone2001}. This is the reason we instead compare the results by numerically obtaining the distribution using the Wishart matrices directly, instead of standard Tracy--Widom distribution.

In Fig. \eqref{fig:twcom} (a), we show the numerically obtained distribution for Wishart matrix of two different dimensions and compare them with the Tracy--Widom distribution, $F_{1}$, the numerically obtained results for infinite dimensional matrices. We observe that while for the Wishart matrices of dimensions $2^{8} \times 2^{8}$ ($W_{8}$), deviations from the $F_{1}$ are observed, these difference vanish when we consider Wishart matrices of size $2^{18} \times 2^{18}$ ($W_{18}$). In Fig. \eqref{fig:twcom} (b), we compare the result of distribution of $\lambda'_{\max}$ for $L=14$ KFIM and COE matrix. The corresponding result of $2^{7} \times 2^{7}$ Wishart matrix, with $10^{6}$ realisations, are also shown. We observe that there is a good agreement between the results of Wishart matrices and the COE matrix, while the KFIM result show clear deviations from either of them. To further compare these three models we do a detailed comparison of mean, variance and skewness in Table \ref{tab:momall}. The result of COE and Wishart matrices show good agreement for all the three cumulants while KFIM differ considerable. It is only for matrices of sizes $2^{18}$ that we observe a good agreement of all the three cumulants. The table thus suggests that it is reasonable for numerical purposes to compare the result with the numerically obtained distribution of Wishart matrices since it is \emph{tending towards Tracy--Widom $F_{1}$ as the dimension increases.}. It is to be noticed that to obtain the eigenvalue statistics of large size Wishart matrices we can use the tri-diagonal form of Laguerre $\beta$ ensemble \cite{dumitriu2002matrix} which allows for efficient sparse matrix evaluations of the \emph{largest} eigenvalue using the Arnoldi method.

\begin{table}[]
\begin{tabular}{|c|c|cccll|lc|llll|}
\hline
\multirow{2}{*}{Moments} & \multirow{2}{*}{TW} & \multicolumn{5}{c|}{Wishart}                                                                                                    & \multicolumn{2}{c|}{COE}                                  & \multicolumn{4}{c|}{KFIM}                                                                            \\ \cline{3-13} 
                         &                     & \multicolumn{1}{l|}{$L= 6$} & \multicolumn{1}{l|}{$L= 7$} & \multicolumn{1}{l|}{$L=8$}  & \multicolumn{1}{l|}{$L=9$}  & $L= 18$ & \multicolumn{1}{l|}{$L=12$} & \multicolumn{1}{l|}{$L=14$} & \multicolumn{1}{l|}{$L=12$}  & \multicolumn{1}{l|}{$L=14$}  & \multicolumn{1}{l|}{$L =16$} & $L =18$ \\ \hline
Mean                     & -1.207              & \multicolumn{1}{c|}{-1.181} & \multicolumn{1}{c|}{-1.192} & \multicolumn{1}{c|}{-1.199} & \multicolumn{1}{l|}{-1.202} & -1.200  & \multicolumn{1}{l|}{-1.182} & -1.182                      & \multicolumn{1}{l|}{-0.2067} & \multicolumn{1}{l|}{-0.2344} & \multicolumn{1}{l|}{-0.3489} & -0.4746 \\ \hline
Variance                 & 1.608               & \multicolumn{1}{c|}{1.240}  & \multicolumn{1}{c|}{1.371}  & \multicolumn{1}{c|}{1.452}  & \multicolumn{1}{l|}{1.506}  & 1.606   & \multicolumn{1}{l|}{1.240}  & 1.374                       & \multicolumn{1}{l|}{2.057}   & \multicolumn{1}{l|}{1.956}   & \multicolumn{1}{l|}{1.840}   & 1.741   \\ \hline
Skewness                 & 0.293               & \multicolumn{1}{c|}{0.4309} & \multicolumn{1}{c|}{0.387}  & \multicolumn{1}{c|}{0.3522} & \multicolumn{1}{l|}{0.3317} & 0.313   & \multicolumn{1}{l|}{0.4227} & 0.3899                      & \multicolumn{1}{l|}{0.5774}  & \multicolumn{1}{l|}{0.4524}  & \multicolumn{1}{l|}{0.3724}  & 0.3257  \\ \hline
\end{tabular}
\caption{Comparison of moments of Tracy--Widom distribution (TW) with matrices belonging to Wishart ensemble, COE and KFIM. The moments of Wishart matrices approach the Tracy--Widom result for large size; $L=18$. The result for COE follow closely the result of Wishart matrices and thus can be expected with confidence to follow the Tracy--Widom result for large matrices. The KFIM result does not match the COE or the Wishart matrix result even for $L=18$ and still show deviations from the Tracy--Widom results.}\label{tab:momall}
\end{table}

\section{Out of time ordered correlator}\label{sec:otoc}
\begin{figure}[ht]
    \centering
    \includegraphics[width=  0.5\linewidth]{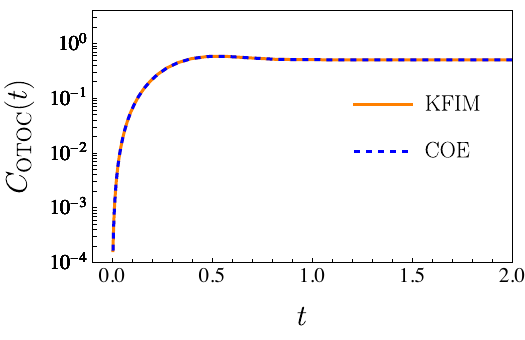}
    \caption{Growth of OTOC with time for KFIM and COE random matrix. The results for the two overlap each other thus showing that KFIM follows the COE random matrix behavior quite well.}\label{fig:otoc}
\end{figure}

Apart from RMT features there are various other probes of quantum chaos. One of them is the out of time ordered correlator (OTOC) \cite{larkin1969quasiclassical,Maldacena:2015waa,Hashimoto:2017oit}. It is typically defined as 
\begin{equation}
    C_{\mathrm{OTOC}}(t)= \braket{[\mathcal{O}_{1}(t),\mathcal{O}_{2}(t=0)]^{2}},
\end{equation}
where $[\cdot]$ denotes the commutator and $\mathcal{O}_{1}(t), \mathcal{O}_{2}(t)$ are the operators at time $t$. For numerical purposes we take $\mathcal{O}_{1}= \mathcal{O}_{2}= \sigma^{z}_{L/2}$, $L=12$ and 100 realization each for KFIM and COE. For a chaotic system the generic behavior of the OTOC is that it grows exponentially upto the Ehrenfest time and then saturates to a constant value at late times. The results for KFIM and COE follow this generic behavior quite well and the corresponding results are shown in Fig. \ref{fig:otoc}. To be noticed that the Ehrenfest time here is extremely small as is evident from the extremely quick approach of OTOC to its saturation values. It can also be observed that for both models the results agree quite well and the corresponding numerical curves overlap. This thus confirms that the measures like OTOC do not really capture the effect of differences from RMT that might occur in the distribution of maximum Schmidt coefficients.

\section{Polynomial Filtering}
Usually for finding the eigenvalues and corresponding eigenvector of large matrices in the mid of the spectrum shift-invert method has been quite successful. However, it has the disadvantage that the memory requirements (and its scaling with size of the matrix) are huge and this thus restrict the method to not very large system sizes. To circumvent this problem for the unitary matrices a polynomial filtered exact diagonalization (POLFED) was proposed in \cite{Luitz_2021,Sierant:2022xtl}. This method has the advantage that the memory requirements are less as compared to shift-invert method for the same-size matrix eigenvalue problem. In our present work we use it for KFIM with $L= 13, 14, 15,16, 18$. For a unitary operator $U$ filtering is done using the following polynomial:
\begin{equation}
    g(U)= \sum_{m=0}^{\kappa} e^{-i \phi} U^{m},
\end{equation}
where we take $ \kappa=\lf 0.8 \frac{2^{L+1}}{2^{L/2+1}}\rf $. We find that with Krylov dimension of $\dd_{K}= \lf  2^{L/2+2}\rf $ we obtain $\dd_{K}/2$ good eigenvalues and eigenvectors. 
Although we aimed to use all the $\dd_{K}/2$ eigenvectors, we found that for a few of the realizations typically last 2 eigenvectors has residue $\sim 10^{-8}$, so we omitted them for the calculation purposes. To quantify the goodness of the Krylov vectors and eigenvalues obtained we calculate the residue. If $U_{A}$ is the Hessenberg matrix corresponding to the unitary matrix $U$ obtained using Arnoldi iteration, and let $(\lambda_{i},V_{i})$ be the eigenvalue and the corresponding  eigenvector respectively of $U_{A}$. Then residue is defined as
\begin{equation}
    r_{i}= ||U_{A}.V_{i} - \lambda_{i} V_{i}||, 
\end{equation}
where $||\cdot||$ denotes the Euclidean norm of the vector.

We make use of the matrix-free implementation of Arnoldi in Mathematica as can be found in \cite{matrixfreemma}. The maximum residue, $\mathrm{Max}(r_{i})$, that we obtained for various $L$ are shown in Table \ref{tab:residuekfim}. 

\begin{table}[h]
    \centering
    \begin{tabular}{|c|c|}
    \hline
         $L$&$\mathrm{Max}(r_{i})$ \\\hline
       13 & $2.83 \times 10^{-15}$ \\\hline
       14 & $2.03 \times 10^{-15}$ \\\hline
        15 & $2.37 \times 10^{-15}$ \\\hline
         16 & $2.54 \times 10^{-15}$ \\\hline
         18 & $3.46 \times 10^{-15}$ \\\hline
    \end{tabular}
    \caption{Maximum residue for various values of $L$ for KFIM.}\label{tab:residuekfim}
\end{table}

Since the Arnoldi method requires only the information of the matrix-vector product, this allows us to simplify the calculation for the KFIM further by using the strategy outlined in \cite{Sierant:2022xtl}. We assume that the vector $v$ used in the recursion is in the $\sigma^{z}$ basis. To convert it into the $\sigma^{x}$ we can make use of fast Walsh–Hadamard transform. In this way while working with $U_{\kfim}.v$ we can work effectively with diagonal $U_{\kfim}= e^{-i H_{z}}. e^{-i H_{x}}$, where each of the matrix is diagonal and the basis of the vector is change using Walsh-Hadamard transform before and after multiplying with the $e^{-i H_{x}}$ part.

\end{document}